\documentclass[10pt,preprint]{aastex}

\def \Halpha{{H$\alpha$\ }}
\def \eg{{e.g.,}}

\def \etal{{et~al.\null}}
\def \ie{{i.e.,}}
\def \vs{{vs.\null}}
\def\kms{{km~s$^{-1}$}}

\shorttitle{Planetary Nebulae in M33}

\shortauthors{Ciardullo et al.}

\begin{document}

\title{The Planetary Nebula System of M33}

\author{Robin Ciardullo\altaffilmark{1}, Patrick R. Durrell, Mary Beth
Laychak\altaffilmark{1}, Kimberly A. Herrmann, Kenneth Moody}
\affil{Department of Astronomy \& Astrophysics, The Pennsylvania State
University
\\ 525 Davey Lab, University Park, PA 16802}
\email{rbc@astro.psu.edu, pdurrell@astro.psu.edu, mary@cfht.hawaii.edu,
herrmann@astro.psu.edu, moody@astro.psu.edu}

\author{George H. Jacoby}
\affil{WIYN Observatory, P.O. Box 26732, Tucson, AZ 85726}

\email{jacoby@wiyn.org}

\and

\author{John J. Feldmeier\altaffilmark{2}}
\affil{Department of Astronomy, Case Western Reserve University \\10900 Euclid
Ave., Cleveland, OH 44106-1712}
\email{johnf@bottom.astr.cwru.edu}

\altaffiltext{1} {Visiting Astronomer, Kitt Peak National Observatory,
National Optical Astronomy Observatory, which is operated by the Association
of Universities for Research in Astronomy, Inc. (AURA) under cooperative
agreement with the National Science Foundation.}

\altaffiltext{2} {NSF Astronomy and Astrophysics Postdoctoral Fellow}

\begin{abstract}
We report the results of a photometric and spectroscopic survey for 
planetary nebulae (PNe) over the entire body of the Local Group spiral 
galaxy M33.  We use our sample of 152~PNe to show that the bright end 
of the galaxy's [O~III] $\lambda 5007$ planetary nebula luminosity 
function (PNLF) has the same sharp cutoff seen in other galaxies.  
The apparent magnitude of this cutoff, along with the DIRBE/IRAS 
foreground extinction estimate of $E(B-V) = 0.041$, implies a distance 
modulus for the galaxy of $(m-M)_0 = 24.86^{+0.07}_{-0.11}$ 
($0.94^{+0.03}_{-0.05}$~Mpc).  Although this value is $\sim 15\%$
larger than the galaxy's Cepheid distance, the discrepancy likely 
arises from differing assumptions about the system's internal
extinction.  Our photometry, which extends more than 3~mag down the PNLF,
also reveals that the faint-end of M33's PN luminosity function is 
non-monotonic, with an inflection point $\sim 2$~mag below the PNLF's 
bright limit.  We argue that this feature is due to the galaxy's large 
population of high core-mass planetaries, and that its amplitude may 
eventually be a useful diagnostic for studies of stellar populations.

Fiber-coupled spectroscopy of 140 of the PN candidates confirms that M33's PN
population rotates along with the old disk, with a small asymmetric drift of 
$\sim 10$~\kms.  Remarkably, the population's line-of-sight velocity dispersion
varies little over $\sim 4$~optical disk scale lengths, with $\sigma_{\rm rad}
\sim 20$~\kms.  We show that this is due to a combination of factors, 
including a decline in the radial component of the velocity ellipsoid at 
small galactocentric radii, and a gradient in the ratio of the vertical 
to radial velocity dispersion.  We use our data to derive the dynamical 
scale length of M33's disk, and the disk's mass-to-light ratio.  Our most 
likely solution suggests that the surface mass density of M33's disk 
decreases exponentially, but with a scale length that is $\sim 2.3$
times larger than that of the system's IR luminosity.  The large scale length 
also implies that the disk's $V$-band mass-to-light ratio changes from 
$M/L_V \sim 0.3$ in the galaxy's inner regions to $M/L_V \sim 2.0$ at 
$\sim 9$~kpc.  Models in which the dark matter is distributed in the 
plane of the galaxy are excluded by our data.

\end{abstract}


\section{Introduction}
For many years, H~I and optical rotation curves have been used to establish the 
presence of dark matter halos around spiral galaxies \citep[\eg][]{fg79,
ashman, combes}.  However, the mass profiles of these halos are still in doubt:
from the rotation curves alone, it is impossible to decouple the gravitational
contribution of the galaxy's visible disk mass from that of its dark halo
\citep[\eg\ see][and references therein]{rubin}.  Most investigations
begin with the assumption that a disk's mass-to-light ratio is constant
with radius \citep[\eg][]{kent86, palunas, sofue}, but evidence in support
of this hypothesis is limited.  In fact, independent estimates of disk 
mass exist for only a handful of spirals \citep[\eg][]{bottema, n5170, 
ic5249, n488, n2985}, and in these systems, the data are restricted to the 
galaxies' inner $\sim 2$~scale lengths.  Thus, while the information to date 
is consistent with the constant mass-to-light ratio hypothesis, the 
applicability of these studies to the dark-matter-dominated outer regions of 
spirals is limited.

The best way to independently determine the mass within a spiral's disk is
through the motions of its old disk stars.  However, the traditional approach
to doing this via absorption-line spectroscopy is extremely challenging.
Because the surface brightness of spiral galaxies declines exponentially
with radius, observations at large radii are quite demanding and require
intricate procedures for data reduction.  This limits the effectiveness
of the technique to the inner one or two disk scale lengths.  Fortunately,
there is an alternative.   Instead of measuring the Doppler widths of 
integrated starlight, one can derive velocity dispersions directly from the 
motions of individual stars.  Planetary nebulae (PNe) are the ideal targets 
for this purpose.  PNe are excellent tracers of the galactic light \citep{p2}, 
easily identifiable via their strong [O~III] $\lambda 5007$ emission and 
distinctive $I(5007)/I({\rm H}\alpha)$ ratio \citep{p12}, and measurable to a 
precision of a few \kms\ via medium dispersion spectrographs.  Moreover, 
because the progenitors of PNe are intermediate and low mass stars, their 
velocity dispersion is representative of the old components of a galactic disk.

Here we describe a photometric and spectroscopic survey of the planetary
nebula system of the Local Group spiral galaxy M33  (NGC~598). M33 is an
excellent target for such an investigation.  First, the galaxy is nearby, 
making large numbers of planetaries available for study.  Second, since
M33's photometric structure is well known at both optical \citep{deV59, 
guidoni, kent} and infrared \citep{rv94, hipp} wavelengths, information is 
available on the distribution of both the young and old stars of the system. 
Finally, M33's H~I rotation curve is well-measured out to $\sim 20$~kpc (\ie\
$\gtrsim 10$~disk scale lengths).  Consequently, detailed models exist for 
the orientation of the galactic disk and the system's total potential
\citep{rwl76, cs97, cs00}.  

We begin our study by describing our photometric survey of the galaxy, and 
comparing our [O~III] $\lambda 5007$ and H$\alpha$ PN measurements to those
of \citet{magrini33a, magrini33b}.  We demonstrate that there is generally
good agreement between the samples, although our [O~III] $\lambda 5007$
magnitudes are systematically brighter than the \citet{magrini33a} values by
$\sim 0.12$~mag.  In Section 3, we present M33's [O~III] $\lambda 5007$ 
planetary nebula luminosity function (PNLF) and show that at faint magnitudes,
the function has the same non-monotonic behavior as the PNLF of the Small 
Magellanic Cloud.  We interpret this turnover as a population effect and
speculate on how the shape of the PNLF can be used to probe the mix of
stellar populations within a galaxy.  In Section~4, we use our data to 
derive a new PNLF-based distance to the galaxy, and discuss a possible reason 
for the $2 \, \sigma$ difference between our measurement and that inferred 
from the galaxy's Cepheid variables.

After analyzing the photometric data, we describe our spectroscopic survey
of the galaxy's planetary nebula system.  In Section 5, we detail our 
observing and reduction procedures, and quantify the errors in our velocity
measurements.  In Sections 6 and 7, we use these data to examine the
stellar kinematics of the galaxy.  We present the system's line-of-sight
velocity dispersion, estimate the stellar asymmetric drift, and use the
epicyclic approximation to model the galaxy's velocity ellipsoid.  We show
that the line-of-sight velocity dispersion of M33's planetaries varies very
little with galactic radius, and that this is due to a combination of factors,
including a decline in the system's radial velocity dispersion at small
galactocentric distances and a gradient in the ratio of the system's vertical
to radial velocity dispersion.  We also show that M33's vertical velocity 
dispersion decreases with a scale length that is greater than that of the 
galaxy's infrared light.  In Section 8, we discuss possible systematic
errors which may effect our result, including the variation of the PN scale
height with galactocentric distance.  Finally, we use our results to estimate
the surface mass density of M33's disk, and constrain the radial profile of
the galaxy's dark halo.  

\section{Identification of Planetary Nebulae}
\begin{figure}
\figurenum{1}
\plotone{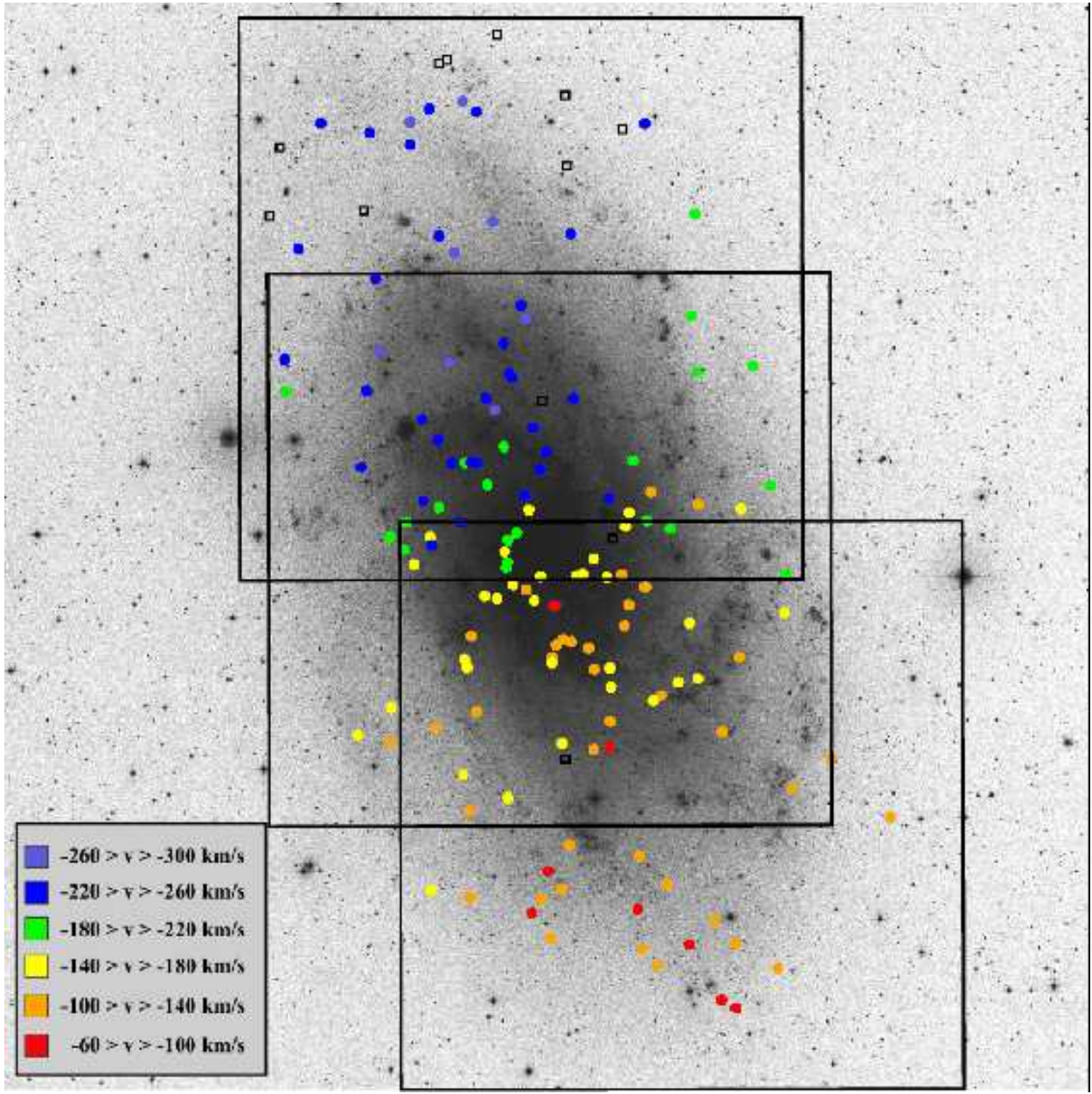}
\figcaption[fig1.eps]{A Digital Sky Survey image of M33.  North is up and east
is to the left; the image is $1\fdg 1$ (18~kpc) on a side. The positions of  
our 152~PN candidates are marked with boxes.  PNe measured spectroscopically
with WIYN are color-coded to show their heliocentric radial velocity.
\label{fig1}}
\end{figure}

M33's PN candidates were identified on Kitt Peak 4~m telescope MOSAIC CCD
images taken as part of the NOAO Local Group Galaxies survey program 
\citep{survey}.  This dataset consists of deep broad- ({\sl UBVRI\/}) and 
narrow- ([O~III], [S~II], and H$\alpha$ + [N~II]) band images of 
10 star-forming galaxies of the Local Group.  Our program made use of the 
survey's M33 frames in [O~III] $\lambda 5007$, H$\alpha$, $B$, and $V$.  These 
data cover a $72\arcmin \times 36\arcmin$ region of the galaxy in three 
overlapping $36\arcmin \times 36\arcmin$ fields, and extend out to a radius 
of $\sim 10$~kpc, or four times the galaxy's disk scale length in blue light 
\citep{deV59}.  Our survey region is shown in Figure~\ref{fig1}; the properties
of the survey frames are summarized in Table~\ref{tab1}.  

To place the \citet{survey} survey frames on a standard system, additional 
exposures in [O~III] $\lambda 5007$ (900~s) and H$\alpha$ (600~s) were taken
with the same telescope/instrument setup on the UT nights of 2001 Dec 22 and 
24.  The seeing during these observations was moderately poor, $1\farcs 4$ in 
[O~III] and $1\farcs 2$ in H$\alpha$, but the sky was photometric, and images
of the \citet{stone} spectrophotometric standards BD+28~4211 and Feige~110
were acquired immediately before and after the exposures.  These supplemental
images were bias-subtracted, flat-fielded, and remapped onto the tangent
plane using the {\tt mscred} routines within IRAF \citep{valdes}.  The latter 
two steps were performed after the iterative subtraction of the telescope's
ghost pupil from the nights' dome flats.

PN candidates were identified by blinking the on-band [O~III] $\lambda 5007$ 
and H$\alpha$ survey images against their $V$-band counterparts.  All spatially
unresolved emission-line objects with little or no continuum flux were 
classified as planetary nebula candidates.  Because medium band-width
continuum images were not part of the \citet{survey} survey, this condition 
effectively meant that all PN candidates had to be either invisible in $V$, or 
have an instrumental $V - \lambda 5007$ color less than $\sim 3$ (the ratio of 
the filters' bandpasses, corrected for transmission differences and the 
inclusion of [O~III] $\lambda 4959$ in the $V$ filter).  Since the seeing on 
the survey frames was generally very good, $\lesssim 1\farcs 0$ ($\sim 4$~pc) 
in [O~III] $\lambda 5007$ and $0\farcs 8$ (or $\sim 3$~pc) in H$\alpha$, this 
simple classification scheme was generally unambiguous.  To confirm the nature 
of the few sources with instrumental colors uncomfortably close to the 
discrimination threshold, \ie\ with $2 < V - \lambda 5007 < 3$, we also 
examined their appearance on the survey's $B$ frames.   Those objects with 
$B$-band fluxes close to that expected from a B main sequence star were 
classified as H~II regions and eliminated from the analysis.  We note that
under these criteria, the two objects closest to our $B$-magnitude threshold
are $\sim 1.5$~mag brighter in $B$ than any other PN in our sample.

Once found, the PN candidates on each [O~III] frame were measured 
astrometrically using a grid of $\sim 200$~reference stars from the USNO-A2.0 
catalog \citep{monet}.  The formal internal errors of our plate solutions 
were $\sim 0\farcs 3$ in both right ascension and declination.  However, 
since M33's nuclear field overlapped the survey's northern and southern
fields by 50\%, almost two-thirds of our PN candidates appeared in two (or
three) frames, and thus could be measured multiple times.  The standard
astrometric error between these measurements, $\sim 0\farcs 15$, is slightly 
better than the internal errors of the individual plate solutions.  This 
suggests that part of $\sim 0\farcs 3$ scatter in the solutions comes from the
proper motions of the reference stars, and that our relative astrometry is 
probably good to better than $\sim 0\farcs 2$.

Relative [O~III] $\lambda 5007$ and H$\alpha$ photometry of the PN candidates
and several bright field stars was accomplished using the {\tt daophot}
point-spread-function fitting (PSF) routines within IRAF \citep{stet87, stet90,
stet92}.  Our photometry was then checked by using {\tt daophot}'s 
{\tt substar} option to subtract off a scaled-PSF representation of
each PN from its position on the frame.  An examination of the residuals
revealed that, while the vast majority of the subtractions were excellent, a 
few objects in or near the galaxy's spiral arms were slightly under- or 
over-subtracted, typically by $\sim 0.1$~mag.  The cause of this error was an 
incorrect estimate of the galactic background, which (particularly in 
H$\alpha$) often included irregular H~II regions, supernova remnants, and 
diffuse knots of emission.  When this occurred, the PN magnitudes were 
manually adjusted until the residuals of the subtractions appeared reasonable.

After determining the raw PN instrumental magnitudes on each frame, the data 
were merged onto a common system by solving the least-squares condition 
required to match the magnitudes of stars in the regions of field overlap 
\citep{cfnjs}.  The instrumental magnitudes were then placed on the standard 
AB system by using the Dec 2001 data to compare large aperture measurements
of field stars to similar measurements made of the \citet{stone}
spectrophotometric standards.  We estimate the error associated with this step 
to be $\sim 0.03$~mag.  Finally, to transform the measured AB magnitudes to 
monochromatic fluxes, we blue-shifted the transmission curves of the 
[O~III] $\lambda 5007$ and H$\alpha$ filters to correct for their temperature 
at the telescope, and applied the photometric procedures for emission-line 
objects described in \citet{jqa} and \citet{p3}.  In doing so, we assumed that
M33's PNe have a mean heliocentric radial velocity of $-180$~\kms\ and a 
dispersion of $\sim 85$~\kms\ about that mean.  Of course, the latter value is 
not strictly applicable to a rotating disk galaxy, but the error introduced
by the peculiar velocity of an individual PNe is never more than 0.01~mag.

Table~\ref{tab2} lists the positions and emission-line strengths of our 152~PN 
candidates.  The [O~III] $\lambda 5007$ magnitudes in the table are related 
to monochromatic flux by
\begin{equation}
m_{5007} = -2.5 \log F_{5007} - 13.74
\end{equation}
where the flux is given in ergs~cm$^{-2}$~s$^{-1}$ \citep{p1}.  Our typical 
photometric errors, as derived internally from {\tt daophot} and externally 
from objects with multiple measurements, are given in Table~\ref{tab3}.

\begin{figure}
\figurenum{2}
\plotone{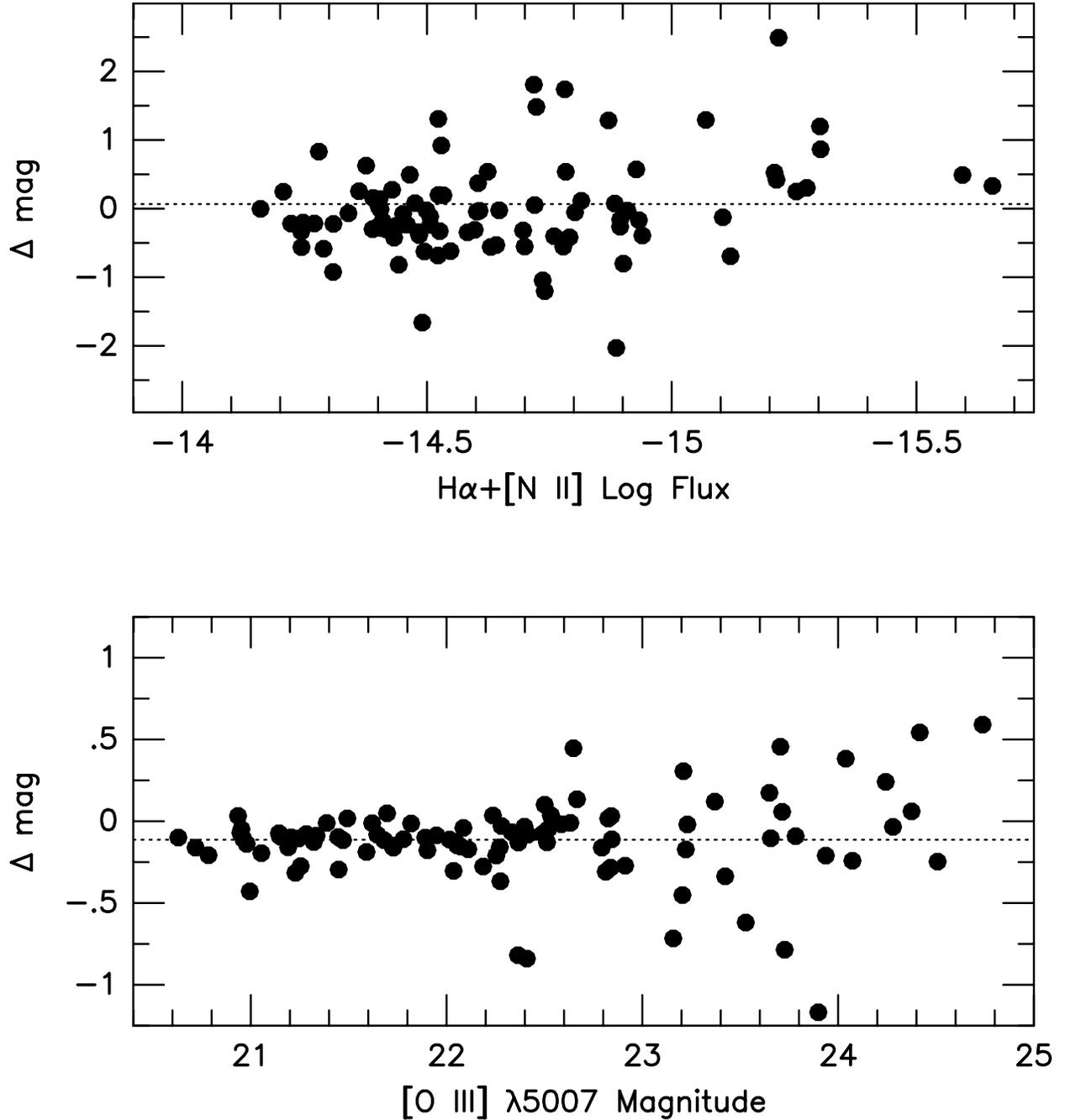}
\figcaption[fig2.eps]{A comparison of our [O~III] $\lambda 5007$ and
H$\alpha$+[N~II] photometric measurements against those of \citet{magrini33b}
for the 101~planetary nebulae common to both surveys.  For the brighter
objects ($m_{5007} \lesssim 23$), the relative [O~III] magnitudes are 
in good agreement, though our measurements are systematically brighter by 
0.12~mag.  The agreement for H$\alpha$ is somewhat poorer:  although our
magnitudes are systematically fainter by only $0.06 \pm 0.05$~mag, the scatter 
for individual measurements is $\sim 0.25$~mag.
\label{fig2}}
\end{figure}

We note that a catalog of M33 PNe has previously been published 
\citep{magrini33a, magrini33b}.  The two datasets agree reasonably well.
Except for a few anomalous outliers, the positions of PNe common to the two 
surveys are virtually identical:  in right ascension and declination, our 
values are systematically larger than the \citet{magrini33b} measurements by 
$0\farcs 07 \pm 0\farcs 06$ and $0\farcs 04 \pm 0\farcs 04$, respectively, and 
the overall dispersion between individual coordinates is $0\farcs 7$. 
Similarly, in H$\alpha$, there is little systematic difference between the
photometry of the two surveys.  On average, our H$\alpha$+[N~II] log fluxes are
$0.06 \pm 0.05$~mag fainter than those of \citet{magrini33b}, and, while the 
dispersion between the measurements is rather large ($\sim 0.25$~mag for 
objects with log fluxes greater than $-15$), this may partly be due to the 
differing contributions of [N~II] within the filters' bandpasses.  In fact,
the only significant difference between the two surveys is in the [O~III] 
$\lambda 5007$ zero point.  Although our [O~III] $\lambda 5007$ magnitudes
agree with those of \citet{magrini33b} in the relative sense (the dispersion 
between the two measurements is $\sim 0.1$~mag for objects brighter than 
$m_{5007} \sim 23$), the zero point of our system is systematically brighter 
by $0.12 \pm 0.05$~mag.  This offset and the dispersion between the 
measurements is shown in Figure~\ref{fig2}.  

Those \citet{magrini33b} PN candidates that were not recovered in our survey
are listed in Table~\ref{tab4}, along with the probable reason for the
discrepancy.  The implied fraction of contaminants ($\sim 22\%$) is consistent
with the 28\% contamination rate estimated by \citet{magrini03} from
follow-up spectrophotometry of the H$\alpha$, [N~II], and [S~II] lines of
36~PN candidates.  However, we note that there are some discrepancies for 
individual objects.  Specifically, \citet{magrini33b} object \#8, which is 
classified by \citet{magrini03} as a planetary nebula, is excluded from our
sample on the basis of its detectable continuum emission and [O~III]
$\lambda 5007$ to H$\alpha$+[N~II] line ratio (see below).  Conversely, 
\citet{magrini33b} objects \#41 and 42 are included in our study, even though 
they are classified by \citet{magrini03} as probable supernova remnants. 

\begin{figure}
\figurenum{3}
\plotone{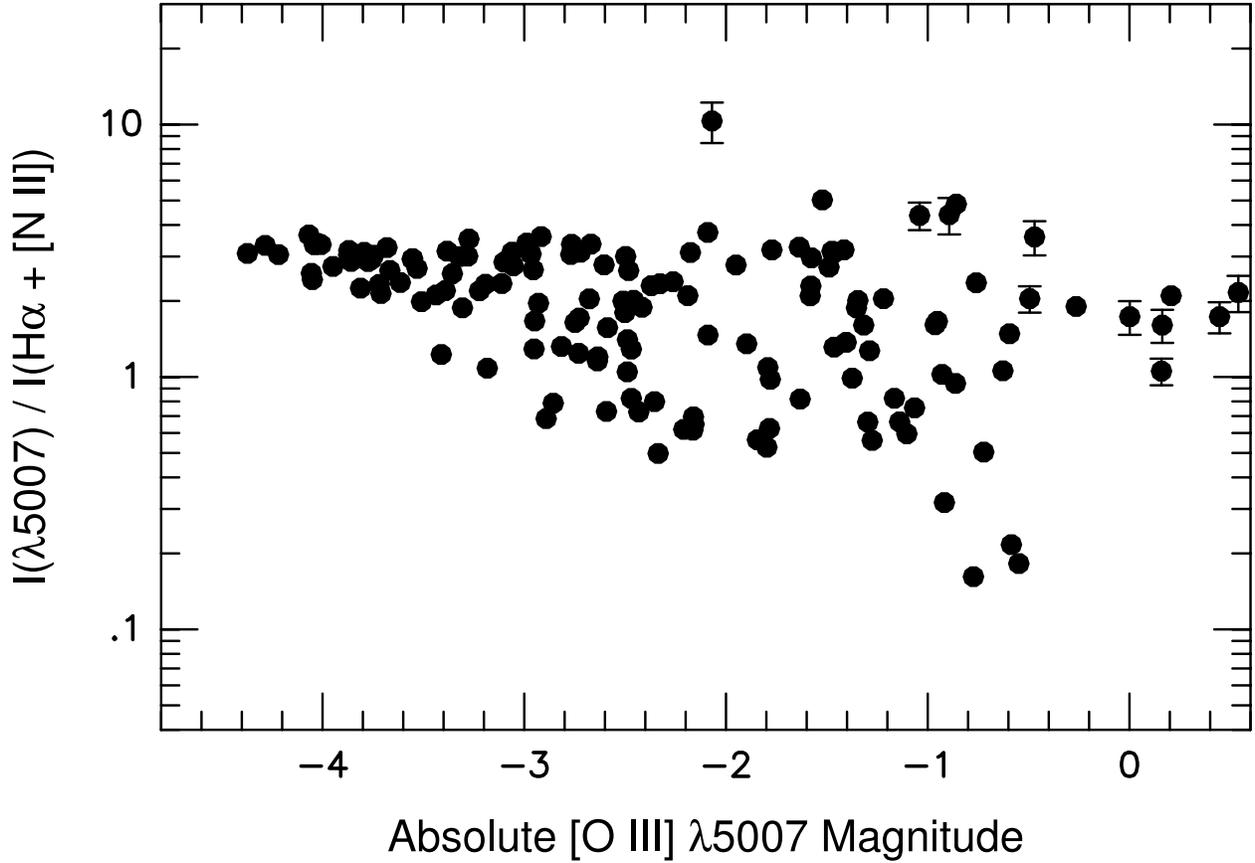}
\figcaption[fig3.eps]{The [O~III] $\lambda 5007$ to H$\alpha$+[N~II] line
ratios for PNe in M33.  The measurements have been corrected for the effects
of foreground Galactic extinction, but not circumstellar extinction or
extinction internal to M33.  For most PNe, the photometric uncertainties
are smaller than the size of the plotted points and have not been displayed.
The region of emission-line space occupied by planetary nebulae has a 
distinctive shape that is the same for all stellar populations.
\label{fig3}}
\end{figure}

Are all the PN candidates listed in Table~\ref{tab2} genuine?  At M33's Cepheid
distance of $\sim 820$~kpc \citep{keyfinal}, our typical seeing of $1\farcs 0$ 
in [O~III] $\lambda 5007$ and $0\farcs 8$ in \Halpha + [N~II] implies that all 
H~II regions or supernova remnants greater than $\sim 3$~pc in size are 
resolved.  This is significantly better than the limit of $\sim 7$~pc 
associated with the poorer seeing exposures of the \citet{magrini33a} survey, 
and as the catalog of \citet{hodge} demonstrates, the density of H~II regions 
in M33 decreases rapidly over this range.  Moreover, since the \citet{survey} 
survey frames reach a limiting magnitude of $V \sim 23.2$, all of M33's stars 
brighter than $M_V \sim -1.3$ are visible via their continuum.  Consequently,
unless an H~II region is excited by a lone main sequence star with spectral
type B3 or later, it will not make it into our sample.  Finally, at the 
bright end of the [O~III] $\lambda 5007$ luminosity function, the excitation
properties of our objects demonstrate that the fraction of contaminants
is negligible.  As pointed out by \citet{p12}, planetary nebulae inhabit
a distinctive cone in [O~III] $\lambda 5007$-H$\alpha$+[N~II] emission-line
space.   While faint PNe (those more than $\sim 3$~mag down the [O~III]
$\lambda 5007$ luminosity function) can have [O~III] $\lambda 5007$ to
\Halpha + [N~II] line ratios anywhere between $0.3 < R < 3$, PNe in the top 
$\sim 1$~mag of the PNLF all have $R \gtrsim 2$.  This contrasts with the 
great majority of compact H~II regions and supernova remnants, which typically 
have $R < 1$ \citep{shaver}.  As Figure~\ref{fig3} illustrates, {\it all\/} of 
our [O~III]-bright PN candidates are high-excitation objects.  Late-type 
B-stars cannot create nebulae with such high [O~III] to H$\alpha$ ratios:
the central stars of all our candidates must be extremely hot, and optically 
faint.  Taken together, the above arguments strongly suggest that the 
contamination fraction in our sample is extremely low.

\section{The Planetary Nebula Luminosity Function}
\begin{figure}
\figurenum{4}
\epsscale{0.9}
\plotone{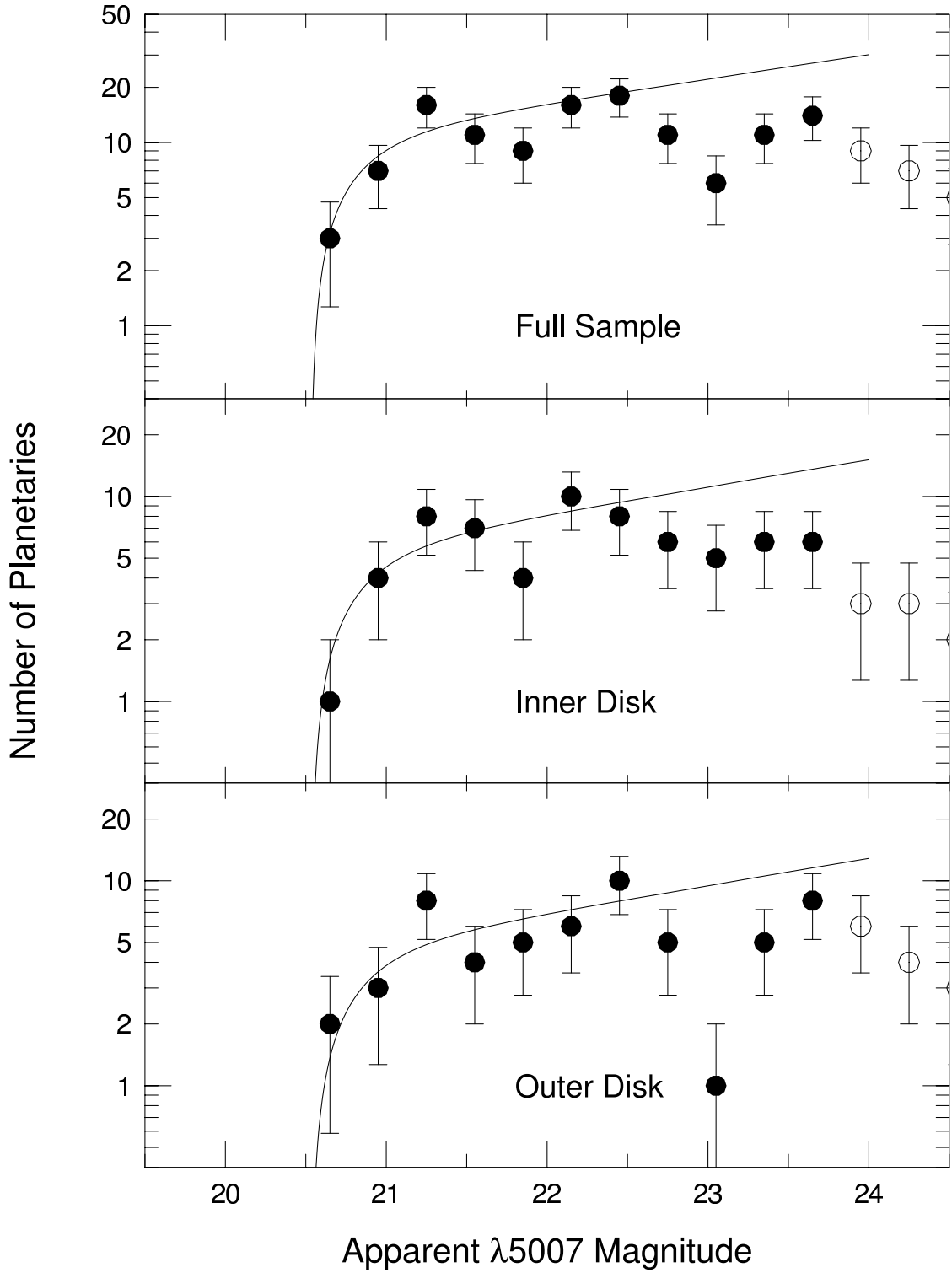}
\figcaption[fig4.eps]{The [O~III] $\lambda 5007$ planetary nebula luminosity
function of M33.   The top panel displays the entire PN sample, the middle 
panel shows only those PNe within $15\arcmin$ of M33's nucleus, and the lower 
panel includes only the PNe outside this isophotal radius.  The curve shows the 
analytic PNLF convolved with the photometric error function and shifted to the 
most-likely distance of the galaxy.  The open circles represent points past the
completeness limit.  Note that the luminosity functions of M33's inner and
outer disk are statistically indistinguishable, and that the PNLF turns over
well before the completeness limit is reached.
\label{fig4}}

\end{figure}

The top panel of Figure~\ref{fig4} displays M33's [O~III] $\lambda 5007$ 
planetary nebula luminosity function.  The sharp cutoff at the bright end of 
the distribution is obvious, as is the function's slow decline at fainter 
magnitudes.  Remarkably, both features are consistent throughout the galaxy.
M33 has a sizable metallicity gradient \citep[$-0.11 \pm 
0.02$~dex~kpc$^{-1}$;][]{garnett}, and the mean oxygen abundance of PNe in
the galaxy's inner disk ($R < 15\arcmin$) is likely to be $\sim 0.5$~dex 
larger than that of PNe outside this radius.  Yet, as the middle and 
lower panels of Figure~\ref{fig4} show, the [O~III] $\lambda 5007$ luminosity 
functions of the two samples are statistically identical.  This constancy 
supports the conclusion of \citet{p12} that the PNLF cutoff is
independent of metallicity for all but the most metal poor systems.  

A second interesting feature exhibited in Figure~\ref{fig4} is the
decline in the PNLF at magnitudes fainter than $m_{5007} \sim 22.5$.  This
roll over is partially real, and partially due to photometric
incompleteness.  To determine the effect of incompleteness, we began by
excluding PNe located within $210\arcsec$ of the galaxy's nucleus; object
detections in this region are difficult due to the bright, irregular
features of the galactic background.  We then measured the local sky 
associated with each remaining object, identified the worst (most uncertain)
background in the sample, and computed the signal-to-noise each PN 
would have if it were projected on that background.  Artificial star
experiments have shown that PN identifications become incomplete when
the signal-to-noise of a detection drops below $\sim 10$ \citep{cfnjs, hui93}.  
Our putative signal-to-noise measurements therefore translate directly into
a completeness limit.  Our completeness limit, $m_{5007} \sim 23.75$, is 
more than a magnitude fainter than the point where the observed PNLF begins 
to decline.

\begin{figure}
\figurenum{5}
\plotone{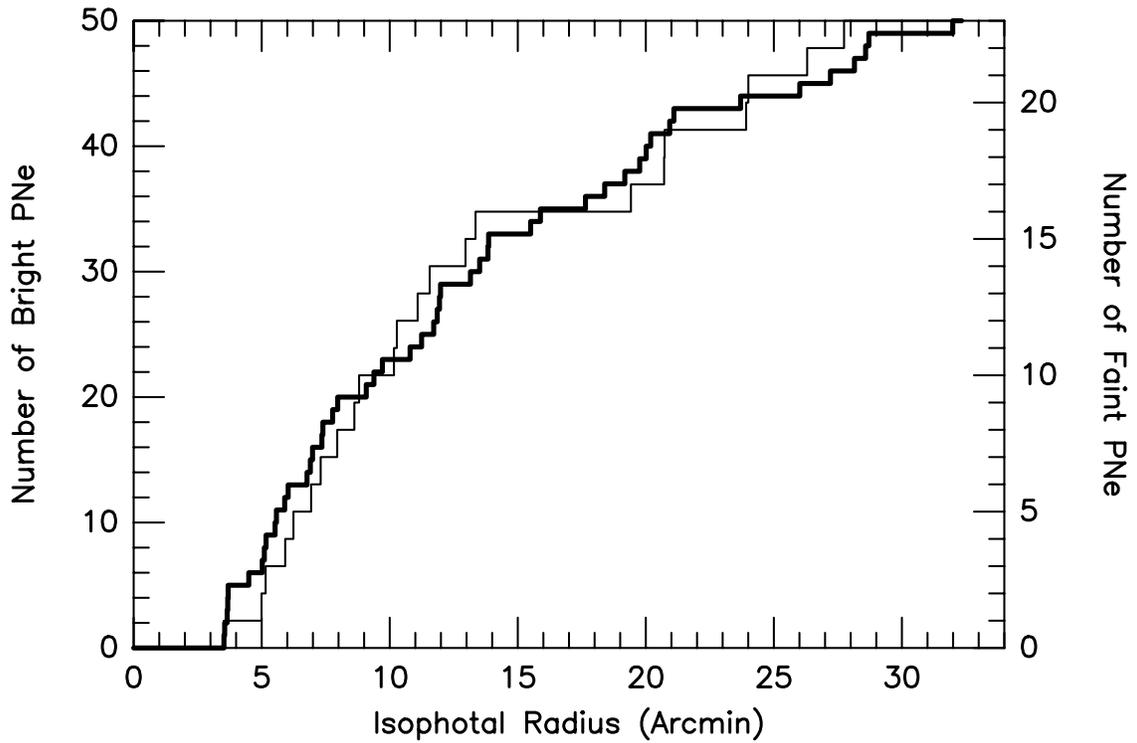}
\figcaption[fig5.eps]{The solid dark line is the cumulative distribution
of isophotal radii for a sample of ``bright PNe'' with $m_{5007} < 22.5$.
The lighter line is a similar distribution for PNe well past the luminosity
function turnover, \ie\ with $23.0 \leq m_{5007} \leq 23.75$.  The two
distributions are statistically identical.  The agreement confirms that 
our PN sample is complete to $m = 23.75$ to within $3\farcm 5$ of the nucleus.
\label{fig5}}
\end{figure}

We can confirm the reality of the dip at $m_{5007} \sim 23$ by comparing 
the radial distribution of ``bright'' PNe ($m_{5007} \leq 22.5$) to a 
similar distribution for PNe with $23.0 \leq m_{5007} \leq 23.75$.  If 
incompleteness were important, then we might expect the latter sample to be 
missing PNe at small radii, where detections are more difficult due to 
the brighter surface brightness of the underlying galaxy.  As Figure~\ref{fig5}
(and the Kolmogorov-Smirnov statistic) show, this is not the case: outside of 
the inner $\sim 210\arcsec$, the cumulative radial distributions for ``bright''
and ``faint'' PNe are identical.  The agreement between the two distributions
demonstrates that our estimate of the detection limit is reasonable and the 
decline in the luminosity function at $m_{5007} \sim 23$ is real.

The decrease in the luminosity function at fainter magnitudes, though
unexpected, is not difficult to explain.  In fact, there are at least two 
possible mechanisms which, in theory, can produce a non-monotonic PNLF{}.  The
first is internal extinction in the host galaxy.  The central extinction in M33
is $A_V \sim 0.9$~mag and there is a significant amount of dust distributed 
throughout the galaxy's disk \citep{rv94, hipp}.   Because the scale height of 
PNe should be larger than that of the interstellar medium \citep{mb81}, one
would expect the bright end of the PNLF to be dominated by objects foreground
to the dust layer.   The observation of bright PNe above the dust lanes of
the edge-on spirals NGC~891 \citep{p7} and NGC~4565 \citep{p10} support this 
idea.  However, at fainter magnitudes, the contribution of extincted PNe may 
conceivably be enough to distort the shape of the PNLF and cause the observed 
luminosity function to turn over.

To investigate this possibility, we modeled the disk of M33 as an isothermal 
sheet of stars (scale height $z_s$) containing an embedded layer of dust 
(scale height $z_d$).  We then adopted the analytic form of the PNLF 
proposed by \citet{p2}
\begin{equation}
N(M) \propto e^{0.307 M} \{ 1 - e^{3 (M^* - M)} \}
\end{equation}
and performed a series of Monte Carlo experiments, in which the ratio of
the two scale heights was allowed to vary between $0.1 < z_d / z_s < 0.5$, 
and the total extinction ranged between $0.3 < A_V < 1.5$.  In no case did 
the mixture of dust and stars produce a non-monotonic PNLF{}.  If the layer of 
dust is thick enough and if the total amount of extinction is large, then both 
the number of PNe and the apparent magnitude of the PNLF cutoff may be affected
by extinction.   However, our analysis suggests that internal extinction, by 
itself, cannot turn the exponentially increasing PNLF of equation (2) into
the peaked luminosity function displayed in Figure~\ref{fig4}.

An alternative explanation for M33's PNLF comes from stellar evolution.  
According to the initial-mass final-mass relation \citep{weidemann}, old
stellar populations produce low-mass central stars with evolutionary
timescales much longer than the timescale for nebular expansion \citep{vw94}. 
To a first approximation, such objects can be modeled as non-evolving 
central stars surrounded by freely expanding gaseous envelopes.  The 
[O~III] $\lambda 5007$ emission of these PNe will decrease with time, $t$,
following the relation
\begin{equation}
F_{5007} \propto N_{\rm O} \, N_e \ V \propto N_e \propto {1 \over R^3} 
\propto {1 \over t^3}
\end{equation}
where $N_{\rm O}$ is the number density of O$^{++}$ ions, $N_e$ is the electron
density, $V$, the volume of the nebula, and $R$, the nebular radius.  For an
ensemble of objects, the number of PNe with [O~III] $\lambda 5007$ magnitudes 
between $M$ and $M + dM$ will then be
\begin{equation}
N(M_{5007}) \propto {dt \over dm} \propto e^{0.307 M}
\end{equation}
This is the exponential law first proposed by \citet{hw63} and used by 
\citet{p2} in their PNLF calculations.  However, it may not be applicable for
younger stellar populations.  In these systems, the typical planetary nebula
will have a higher mass core, and a much shorter timescale for stellar 
evolution.  If this timescale is comparable to, or shorter than, the timescale 
for envelope expansion, then a PN's [O~III] $\lambda 5007$ evolution will be 
driven by the luminosity evolution of its central star, rather than the 
expansion of its nebula.   Since post-AGB stars spend a relatively long
time at high luminosity (as they cross they HR diagram at the end of
shell burning), a short time at intermediate luminosity (when nuclear
reactions stop), and a long time at low luminosity (as their cooling rate 
slows), the PNLF of high-mass objects will be double-peaked, and similar
to that calculated for post-AGB stars by \citet{vw94}.

\begin{figure}
\figurenum{6}
\epsscale{0.9}
\plotone{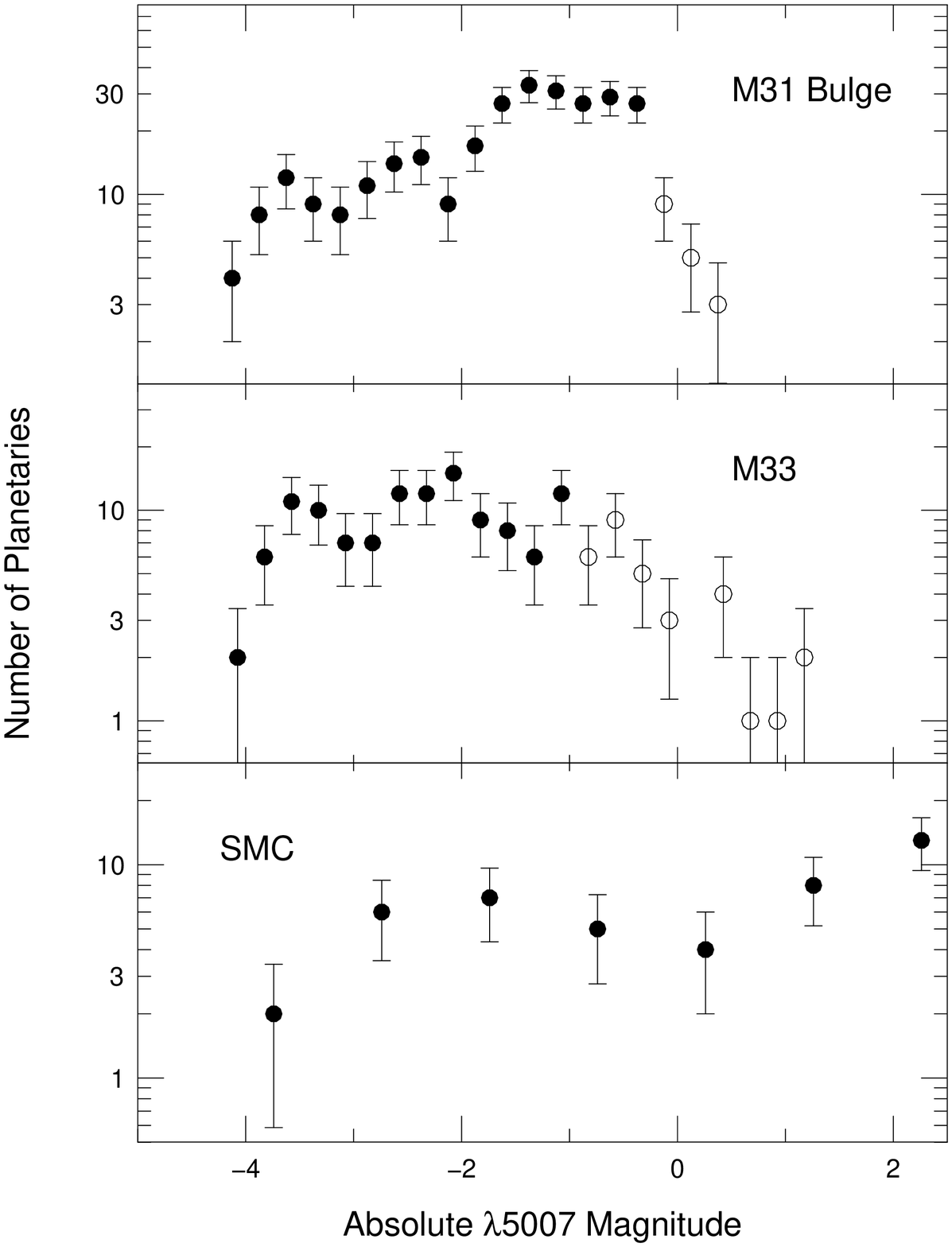}
\figcaption[fig6.eps]{The [O~III] $\lambda 5007$ planetary nebula
luminosity functions of three Local Group stellar populations.  The data for
M31's bulge are taken from \citet{p12}; the SMC luminosity function comes from
\citet{jd02}.  The absolute magnitudes are derived using the Cepheid distances 
to the galaxies \citep{keyfinal, udalski99} and the Galactic extinctions
given by \citet{schlegel}.  While the PNLF of M31's bulge is a monotonically 
increasing function, the luminosity functions of the two star forming systems 
decline at fainter magnitudes.
\label{fig6}}

\end{figure}

Of course, in order to model real systems, one must include central star 
evolution, nebular expansion, and gas dynamics in the analysis, as
well as the mix of stellar populations within a galaxy \citep[see][]{marigo}.
Nevertheless, these simple arguments do explain the different PNLFs found in 
Local Group galaxies.  As Figure~\ref{fig6} demonstrates, the low-mass stars 
of M31's bulge have a PNLF that is well-represented at the faint end by an 
exponential \citep{p12}.  Conversely, M33 and the SMC are actively forming 
stars and their PNLFs are consistent with the bimodal distribution of 
\citet{vw94}.  Given the distinctive nature of the two limiting cases, it 
is possible that the strength of the ``dip'' in the luminosity function may 
someday be a useful probe of stellar populations which are otherwise difficult 
to observe.

\section{The Distance to M33}
Although the shape of the PNLF at fainter magnitudes may change with time, 
the absolute magnitude of the function's bright-end cutoff stays
amazingly constant \citep{mudville, chile}.  Whether this is due to 
convergence in the initial mass-final mass relation \citep{jacoby97},
a fortuitous correlation between central star UV flux and circumstellar
extinction \citep{cj99, chile}, or the contribution of some other
PN-like object \citep{marigo} is unclear.  Nevertheless, the
insensitivity of the PNLF cutoff to population age and metallicity makes
the feature extremely useful as an extragalactic standard candle.

To derive M33's PNLF distance and its formal uncertainty, we followed the
procedures of \citet{p2}.  We took the analytic form of the PNLF given in
equation (2), convolved it with the photometric error vs.~magnitude relation of
Table~\ref{tab3}, and fit the resultant curve to the data via the method of
maximum likelihood.  Since this empirical law assumes a monotonically 
increasing PNLF at faint magnitudes, we limited the range of the fit to
$m_{5007} < 22.5$.  In addition, in order to correct for foreground Galactic 
extinction, we adopted the DIRBE/IRAS-based reddening value of $E(B-V) = 0.041$ 
\citep{schlegel} and used the relation of \citet{ccm} with $R_V = 3.1$ to go 
from differential to total $\lambda 5007$ extinction.  Assuming an absolute 
magnitude for the PNLF cutoff of $M^* = -4.47$ \citep{p12}, the most likely 
distance to M33 is $(m-M)_0 = 24.86$ (0.94~Mpc) with a formal fitting error of 
$+0.05/$$-0.10$~mag.  When we include the systematic errors associated with the 
observation's photometric zero point (0.03~mag), the filter response curve 
(0.03~mag), and the Galactic foreground extinction 
\citep[$0.16 \, E(B-V)$;][]{schlegel}, M33's most likely PNLF distance modulus 
becomes $(m-M)_0 = 24.86^{+0.07}_{-0.11}$, or $0.94^{+0.03}_{-0.05}$~Mpc.

Our PNLF distance modulus of $(m-M)_0 = 24.86^{+0.07}_{-0.11}$ is 0.30~mag
larger than M33's Cepheid distance modulus of $24.56 \pm 0.10$.  If these
uncertainties are accurate, then a simple propagation of errors 
\citep[\ie][]{bevington} implies that the two measurements are discrepant at 
the $\sim 2 \, \sigma$ level.   There are two possible explanations for this 
discrepancy.  

The first involves a metallicity dependence in either (or
both) of the distance indicators.  Thirteen galaxies have both PNLF and
Cepheid distance measurements, and a detailed analysis of these data has 
been performed by \citet{p12}.  Their results are striking:  except for the
most metal-poor systems, the distance residuals are perfectly consistent with
the internal uncertainties of the methods.  Moreover, although a systematic 
shift is seen in low-metallicity ([O/H] $\lesssim -0.5$) systems, the offset is 
exactly that predicted for the PNLF by \citet{djv92}.  This agreement leaves
little room for further metallicity corrections in either the PNLF or
Cepheid methods.  

Nevertheless, since our distance estimate to M33 assumes $M^* = -4.47$, 
it is possible that metallicity does play a role in the distance discrepancy.
However, in order to do so, the mean oxygen abundance of the PNe in our 
sample must be $\sim 0.2$~dex below that of the Large Magellanic Cloud.  
The data of \citet{garnett} show that this is unlikely: only in the 
outermost regions of the galaxy do the H~II region abundances drop to this
level.  Moreover, if $M^*$ were being affected by metallicity, the strong
abundance gradient across the galaxy would cause our measurement of $m^*$
in M33's outer disk to be fainter than that of the inner disk.  As
Figure~\ref{fig4} demonstrates, this is not observed.  In fact, the PNLF
cutoff for the sample of PNe with $R > 15\arcmin$ is marginally brighter
than that for the PNe inside this radius (by $0.15 \pm 0.17$~mag).  This 
constancy strongly suggests that the low metallicity of M33 is not causing 
us to overestimate $M^*$ and the distance to the galaxy.  

As pointed out by \citet{p12}, a more likely explanation for M33's distance
discrepancy involves divergent assumptions about the galaxy's internal 
reddening.  M33's Cepheid distance uses an extinction estimate that is derived 
from multicolor photometry of the Cepheids themselves \citep{freed91, keyfinal},
while the PNLF method assumes the DIRBE/IRAS Galactic value.   The difference 
between these values, $E(B-V) = 0.17$, is one of the largest observed for any 
Cepheid galaxy.  If we were simply to adopt the Cepheid reddening estimate of 
$E(B-V) = 0.21$ \citep[as did][]{magrini33a}, or if the internal reddening
were reduced to one typical of other Cepheid galaxies, the PNLF and Cepheid 
distance indicators would be in much better agreement.  Additional support for 
this hypothesis comes from the galaxy's red giant stars.  The distance moduli 
derived by \citet{kim} using the tip of the red giant branch ($24.81 \pm 0.04$ 
(random) $\pm 0.13$ (systematic)) and the location of the red clump ($24.80 
\pm 0.04$ (random) $\pm 0.05$ (systematic)) are both consistent with the PNLF 
value.

\section{Planetary Nebula Spectroscopy}
The radial velocities of M33's planetary nebulae were measured with the WIYN
telescope, the HYDRA bench spectrograph, and a 600~lines~mm$^{-1}$ grating
blazed at $10\fdg 1$ in first order.  Most of the PNe were observed on the
photometric nights of 2002 Oct 3-5, with the $2\arcsec$ red fiber cable and
a telescope-instrument combination which produced spectra with 2.8~\AA\ 
(168~\kms) resolution and 1.4~\AA~pixel$^{-1}$ dispersion over the wavelength
range between 4500~\AA\ and 7000~\AA.   The data were taken using six fiber
setups, each of which targeted $\sim 50$~PNe with $\sim 5$~fibers devoted to
the sky.  The total exposure time for each setup was $\sim 3.5$~hours, and, 
in order to allow for the removal of cosmic rays and other instrumental
artifacts, the integrations were subdivided into a series of 30~min
exposures.  To provide a check on the repeatability of the measurements,
$\sim 75\%$ of the PNe were observed multiple times, and 7 PNe, ranging
in brightness between $21.0 < m_{5007} < 25.0$ were targeted in four or more
setups.

To supplement these data, additional spectra for 24~PN candidates were
obtained by R. Chandar and H. Ford on the nights of 2003 Jan 3-5 using
HYDRA's $3\arcsec$ blue fiber cable and a 400~lines~mm$^{-1}$ grating blazed 
in first order at $4 \fdg 2$.  These data, which covered the wavelength range
between 3660~\AA\ and 6860~\AA, had a slightly lower dispersion than
the Oct 2002 data ($\sim 6$~\AA\ resolution at 1.56~\AA~pixel$^{-1}$),
but longer  ($\sim 5$~hr) exposure times partially compensated for this
drawback.  These Jan 2003 observations yielded three additional PN
velocities, and reduced the measurement errors of 21 other objects.

Data reduction was accomplished with the {\tt dohydra} task within IRAF{}.
Flat-fielding was performed using dome flat exposures obtained at the 
beginning of each night, and wavelength calibrations were found using a 
series of CuAr comparison arcs taken before, during, and after the science 
exposures of each setup.  The solutions derived from these comparison arcs
had a rms dispersion of $\sim 0.03$~\AA\ ($\sim 2$~\kms) and were stable
throughout each setup.  Next, the spectra were linearized to a common
wavelength scale, and sky subtracted using an average sky spectrum determined
from the setup's sky fibers.  Because our observations focussed on regions
of the spectrum well away from any airglow emission, and were performed during
dark time at moderately high ($R \sim 2000$) dispersion, the details of this 
subtraction make very little difference to the final result.  After
sky subtraction, the individual exposures of each setup were co-added
and shifted into the barycentric rest frame to produce a single summed
spectrum for each object.

PN radial velocities were derived in a two step process.  First we obtained
an interim velocity for each PNe using the centroids of the bright
emission lines of [O~III] $\lambda\lambda 4959,5007$, H$\alpha$, H$\beta$, 
and (when possible) [N~II] $\lambda\lambda 6548,6584$.  We then used
these velocities to create an ultra-high signal-to-noise PN template by 
shifting all of the spectra into the rest frame and co-adding the data. 
Using this template, we then determined our final PN velocities by 
cross-correlating the individual spectra against the template spectrum
with the {\tt xcsao} task of the {\tt rvsao} package \citep{rvsao} of IRAF{}.
Generally speaking, for faint emission-line objects, velocities derived in 
this way are slightly more accurate than velocities found from the weighted 
average of individual emission-line measurements.  In the case of our M33 
survey, where most of the PNe are bright, the improvement to our measurements 
was marginal.  The mean difference between velocities determined via 
cross-correlation and those found by the line-centroiding ({\tt emsao}) task 
within {\tt rvsao} was only $+0.15 \pm 0.34$~\kms.

At this stage, we verified the internal consistency of our velocity 
measurements by cross-correlating the spectra of each individual setup
against the template spectrum, and intercomparing the results for the 113~PNe 
with multiple measurements.  Table~\ref{tab5} summarizes the results of this
analysis by listing the mean velocity offset of each setup with respect
to the others, and the number of velocity pairs used to derive these values.
From the table, it is clear that systematic setup-to-setup velocity
differences are minimal:  the only setup whose offset is not consistent with
zero is that of Setup~4, and its velocities are low by only $\sim 2.9$~\kms. 
Table~\ref{tab5} also lists the standard error between the individual pairs 
of PN measurements.  This error provides an upper limit to the internal
accuracy of our measurements.

Column~10 of Table~\ref{tab2} lists our velocity measurements for M33's PNe.
Column~9 of the table gives the number of fiber setups employed for each
object, excluding the observations of January 2003.  Those PNe targeted in
January are identified in Column~13, and their velocities are the weighted
mean of the results from the two observing runs.  In total, of the 152~PN
candidates identified on our survey frames, 151 were targeted by WIYN, and 
140~were detected via their emission lines.   (Of the 11 objects that were not 
detected, 7 are extremely faint.)  Note that the errors quoted in Column~8 are 
the internal uncertainties of the cross-correlation technique and do not 
include any setup-to-setup systematic uncertainty.  As demonstrated in 
Table~\ref{tab5}, such errors are small, $\lesssim 3$~\kms.

\section{The M33 Planetary Nebula Velocity Field}
The large scale velocity distribution of M33's PNe is illustrated in
Figure~\ref{fig1}.  The most obvious feature displayed in the figure is the
galaxy's rotation.  M33 is inclined $\sim 56^\circ$ to the line-of-sight
\citep[see][and references therein]{zar89}, and this inclination is reflected
in the radial velocities of the planetaries.  In fact, since M33's 
spheroid-to-disk ratio is extremely small, $S/D \sim 0.02$ 
\citep{bothun, rv94}, all but $\sim 2$ PNe should be rotating along with the 
galactic disk.

To explore the dynamics of M33's PN population, we began by assuming that, 
in the mean, the system's PNe are confined to the disk and move in circular
orbits about the galactic nucleus, which is at $\alpha(2000)$ = 1:33:50.915, 
$\delta(2000)$ = +30:39:36.79 \citep{massey96}.  We then defined 
the inclination and orientation of this disk.  On average, M33 is inclined $i 
\sim 56^\circ$ to the line-of-sight and has a major-axis position angle of 
$\theta = 23^\circ$ \citep{kent, zar89}; these values are consistent with the 
isophotal contours of the inner parts of the galaxy \citep{kent}, as well as 
the kinematics of the system's H~II regions \citep{zar89}.  However, such a 
model is probably too simplistic.  The kinematics of M33's H~I gas 
demonstrates the presence of a significant warp in the galaxy's outer regions 
\citep{rwl76, cs97, cs00}.  This warping is also confirmed by the asymmetrical 
shape of the galaxy's spiral arms \citep{sh80}.  To include this effect, we 
therefore adopted the model of \citet{cs00}, in which the inclination and 
position angle of M33's disk varies with galactocentric distance.  Fortunately,
our analysis depends very little on the details of this assumption:  over the
inner $\sim 30\arcmin$ of the galaxy (\ie\ over most of our survey region),
the difference between the \citet{cs00} model and the simple $i = 56^\circ$ 
$\theta =  23^\circ$ geometry is slight (no more than $4^\circ$ in inclination 
and $5^\circ$ in position angle).  Nevertheless, since M33's warp does become
important outside this range, and since a few of our PNe do have large
isophotal radii, we prefer this approach over the less sophisticated flat-disk
assumption.

After defining M33's inclination and position angle, we 
deprojected the position of each PN to determine its galactocentric radius 
($R$) and position angle with respect to the galaxy's major axis ($\phi$). 
The rotation speeds of the PNe, $V_*$, were then inferred from the 
line-of-sight radial velocities, $V_{\rm rad}$, via 
\begin{equation}
V_* = {(V_{\rm rad} - V_{\rm sys}) \over \sin i \, \cos \phi}
\end{equation}
For the systematic velocity, we used $V_{\rm sys} = -180$~\kms; this value
is based on both radio \citep{RC3} and optical \citep{hvg99} observations
of the galaxy, and is comfortably close to the mean radial velocity 
of $\langle V_{\rm rad} \rangle = -176$~\kms\ found from the PNe of this
study.

The top panel of Figure~\ref{fig7} compares our PN rotational velocities with
the rotation curve of M33's atomic (H~I) and molecular (CO) gas \citep{c03}.
All objects except 14 PNe within $10^\circ$ of the minor axis have been 
plotted.  Overall, the agreement is quite good.  The data display some amount 
of scatter, but this is due to the peculiar radial, tangential, and vertical
velocities of the PNe.  When these non-circular motions are amplified by the 
$\cos \phi$ division, the result is the dispersion seen in the figure.  
Also apparent is the small, but significant, offset between the mean circular
velocity of the PNe and the rotation speed of the gas.  The bottom panel
of the figure shows this offset more clearly by binning the data by
galactocentric radius with 18 objects per bin.  The amplitude of the velocity
lag, or asymmetric drift, is $\sim 10$~\kms, less than half the drift velocity
observed for old-disk stars in the vicinity of the Sun \citep{ku97, db98}.  
However, since the ratio of M33's drift speed to rotation velocity 
is approximately the same as that for the solar neighborhood, our measurement 
suggests that the dynamical state of M33's disk is not too dissimilar to that 
of the Milky Way.  We will return to the issue of asymmetric drift in 
Section~7, where we use it to help constrain the shape of the system's velocity 
ellipsoid.

Another way to examine the motion of M33's PNe is to plot their line-of-sight
velocities after removing the contribution of the rotating stellar disk
(defined as the H~I + CO gas velocity minus the asymmetric drift).  This is 
done in Figure~\ref{fig8}; the top panel of the figure shows the individual 
PN velocities, while the lower panel bins the velocities to show the 
line-of-sight velocity dispersion.  The figure again demonstrates that the 
non-circular motions of the PNe are significant, with $\sigma_{\rm rad} \sim 
25$~\kms.  This is more than twice the $9 \pm 4$~\kms\ dispersion found for 
the galaxy's H~II regions \citep{brandt65, zar89}.  In addition, our innermost 
measurement of $\sigma_{\rm rad} = 22^{+5}_{-3}$~\kms\ is very close to the 
galaxy's central value of $24.0 \pm 1.2$~\kms\ found from absorption line
spectroscopy \citep{gebhardt}.  Since M33 shows no dynamical evidence for a
central black hole or mass concentration, this agreement confirms that
we are, indeed, measuring a velocity dispersion that is representative of
old disk stars.

Figure~\ref{fig8} also supports our contention that there are very few non-disk
PNe in our sample.  Of the 140 PNe surveyed, only two objects, PN 67 and PN 24, 
have velocities that are more than $2.3~\sigma$ from the galactic rotation.
These planetaries, which deviate by 2.8 and $3.0 \, \sigma$ respectively, 
may {\it possibly\/} belong to M33's spheroidal component.  On the other
hand, in a normally distributed population of 140 PNe, we should expect to 
find $\sim 2$ objects with velocities more than $\sim 2.5 \, \sigma$ away 
from the mean.  Thus, it is possible that {\it none\/} of our PNe belong
to M33's halo, and all are rotating along with the disk.  This confirms 
the photometric results of \citet{bothun}, \citet{rv94}, and \citet{tiede} that
M33's spheroidal component is negligible compared to its disk.

The most surprising aspect of Figure~\ref{fig8} is how little the line-of-sight
velocity dispersion depends on galactocentric distance.   In the $K$-band,
where internal extinction presumably is unimportant, M33's disk scale length
is $\sim 5\farcm 8$ \citep{rv94}.   Our observations, which extend out to 
a distance of $34\arcmin$, therefore sample $\sim 6$~scale lengths of the
galaxy.  If M33's disk is isothermal, then the galaxy's vertical velocity 
dispersion should decline significantly over this range.  Specifically, if 
$z_0$ is the scale height of the planetaries and $\Sigma$ the disk-mass 
surface density, then under the isothermal disk approximation
\begin{equation}
\sigma_z^2(R) = \pi G \, \Sigma(R) z_0 
\end{equation}
Since M33's stellar and gas surface density both decline exponentially
\citep{guidoni, rv94, c03}, it is reasonable to expect $\Sigma$, and therefore 
$\sigma_z^2$, to do the same.  Unless the shape of the galaxy's velocity 
ellipsoid changes dramatically, or M33's disk is significantly flared, 
$\sigma_{\rm rad}$ should decrease by more than a factor of 20 over our
survey region.  Figure~\ref{fig8} clearly demonstrates that it 
does not.

\section{Modeling the Velocity Ellipsoid}
Because the disk of M33 is inclined $\sim 56^\circ$ to the line-of-sight, 
peculiar stellar motions in the radial ($\sigma_R$), tangential 
($\sigma_{\phi}$), and vertical ($\sigma_z$) directions all contribute to the 
scatter seen in Figure~\ref{fig8}.  Specifically, the line-of-sight 
velocity dispersion is related to the galaxy's velocity ellipsoid through
\begin{equation}
\sigma_{\rm rad}^2 = \sigma_R^2 \sin^2 \phi \, \sin^2 i + 
\sigma_{\phi}^2 \cos^2 \phi \, \sin^2 i + \sigma_z^2 \cos^2 i
\end{equation}
Consequently, in order to interpret the figure, the shape 
of M33's velocity ellipsoid needs to be constrained.

To do this, we began by using the epicyclic approximation to write 
$\sigma_{\phi}$ in terms of $\sigma_R$ 
\begin{equation}
\sigma_{\phi}^2 =  \sigma_R^2 \left( {1 \over 2} + 
{1 \over 2} \, {\partial \ln V_c \over \partial \ln R} \right)
\end{equation}
\citep{bt}, where the circular rotation velocity, $V_c$, is known from the
motion of the system's H~I gas.  We then parameterized equation (7) in terms 
of the two remaining unknowns, $\sigma_z$ and $\sigma_R$, and, for each radial
bin, we computed the relative probability of observing the given set of PN 
velocities as a function of these two variables.  Once these probabilities
were established, we normalized their values to one to generate 
the likelihood of each solution.

Note that this type of calculation requires that limits be placed on the
variables, so that outside the range of analysis, the probability of a 
solution is identically zero.  For $\sigma_z$, our analysis included all
values between $0 < \sigma_z < 100$~\kms; given the $\sim 100$~\kms\
rotation speed of the galaxy, the true value of $\sigma_z$ certainly 
falls within this range.  Our choice of limits for $\sigma_R$ was more 
difficult.  In the vicinity of the Sun, the ratio of the vertical to radial 
velocity dispersion for old disk stars is $\sigma_z / \sigma_R \sim 0.5$ 
\citep{bienayme}, and surveys of other spirals consistently find ratios in 
the range $0.3 < \sigma_z / \sigma_R < 0.9$ \citep[\eg][]{n488, n2985, bottema,
vdKdG}.  Unfortunately, almost all these data are for stars between
$\sim 1$ and $\sim 2$~disk scale lengths from the galactic center:  there are
no measurements of $\sigma_R$ or $\sigma_z$ over the entire range of radii 
considered in this paper.  Thus, our only guidance on the shape of M33's 
velocity ellipsoid comes from numerical simulations.  According to the models 
of \citet{villumsen} and \citet{jb90}, disk heating by molecular clouds and
spiral structure should drive an initially isotropic velocity dispersion
towards values of $\sigma_z / \sigma_R$ as low as $\sim 0.4$.  For a lower
limit on $\sigma_R$, we can therefore demand that $\sigma_R > \sigma_z$.  For
the upper limit on this quantity, we can use the requirement that M33's disk
be stable against buckling instabilities, which occur when
$\sigma_z / \sigma_R \lesssim 0.3$ \citep{toomre66, araki, ms94}.  To be
slightly conservative, we therefore limited our analysis to 
$0.25 < \sigma_z / \sigma_R < 1.0$.

Figure~\ref{fig9} displays the likelihood contours produced by our data.
As is obvious from the figure, $\sigma_R$ is better constrained
than $\sigma_z$.  This is understandable: not only did our boundary
conditions ensure that $\sigma_R > \sigma_z$, but the former variable
also receives the contribution of $\sigma_{\phi}$, via the epicyclic 
approximation.   As a result, the line-of-sight velocity dispersion 
largely reflects the behavior of $\sigma_R$; our limits on $\sigma_z$
come primarily from the constraints placed on the shape of the velocity
ellipsoid, and secondarily from the fits.

The more important property shown in Figure~\ref{fig9} is the absence of
an obvious radial gradient.  There is very little difference between the 
fits for PNe near the center of M33 and PNe at $\gtrsim 4$~disk scale 
lengths.  If the shape of the galaxy's velocity ellipsoid is roughly 
constant with radius, then the data imply that $\sigma_z$ must also be 
very nearly constant.  On the other hand, if the vertical velocity 
dispersion truly declines with radius (as it must in a constant 
mass-to-light ratio disk), the dispersion ratio $\sigma_z / \sigma_R$ 
must also decline.

To make further progress, we can apply two additional constraints to our
data.  The first is the requirement that M33's disk be stable.  According
to \citet{toomre}, in order for a thin stellar disk to be stable against
axisymmetric perturbations, its radial velocity dispersion must satisfy
the condition
\begin{equation}
\sigma_R > {3.36 \, G \ \Sigma \over \kappa}
\end{equation}
where the epicyclic frequency, $\kappa$, is obtainable from the rotation curve 
via
\begin{equation}
\kappa = {V_c \over R} \left( 2 + 2 \, {\partial\ln V_c \over 
\partial\ln R} \right)^{1/2} 
\end{equation}
\citep{bt}.  In addition, for stability against non-axisymmetric perturbations
(\ie\ bar formation), the disk must also satisfy the (slightly more stringent)
condition
\begin{equation}
\sigma_R > {3.36 \, G \ \Sigma \over \kappa} \cdot {2 V_c \over R \, \kappa}
\end{equation}
\citep{moro1, moro2, moro3}.  If we combine this equation with the isothermal
disk approximation given by equation (6), we obtain a constraining relationship
between the vertical and radial velocity dispersions
\begin{equation}
\sigma_z < \kappa \left( {\pi z_0 R \, \sigma_R \over 6.72 \, V_c} \right)^{1/2}
\end{equation}

To use this relation, we need to know $z_0$, the scale height of PNe in M33's
disk.  Our probability contours of Figure~\ref{fig9}, in combination with
the buckling instability condition, allow us to exclude values of $z_0 \lesssim 
90$~pc in M33's outer regions ($R > 6$~kpc).  To improve upon this number,
however, we must turn to other galaxies.  In the Milky Way, measurements of
$z_0$ are notoriously uncertain, with published values ranging all the way from
90~pc to 300~pc \citep[\eg][]{pot84, maciel, zp91, cs95, phillips}. 
Unfortunately, in other galaxies the situation is worse: internal extinction
in the planes of edge-on galaxies make reliable measurements of the
PN distribution nearly impossible \citep[see][]{p7, p10}.  For lack of a 
better alternative, we were therefore forced to adopt an intermediate value
of $z_0 \sim 175$~pc as the most-likely scale height of our test particles.
This value has two justifications.   First, a Milky Way scale height of $z_0
\sim 175$~pc implies a vertical PN distribution similar to that of Galactic
main-sequence F-stars \citep{mb81} and a total PN population that is consistent
with PN observations in other galaxies \citep{peimbert}.  Second, \citet{bm02}
have found a correlation between the infrared ($K_s$) central surface
brightness of a galaxy and the ratio between the galaxy's vertical and radial
scale lengths. If one applies this relation to M33 \citep{rv94} and normalizes
the result using the ratio of our Galaxy's PN scale height to the total ``old
thin disk'' scale height \citep{chen}, then $z_0$ again becomes $\sim 175$~pc.
Obviously, this value carries a substantial uncertainty of at least
$\sim 25\%$.

We note that, in theory, the scale height of a galactic disk may change with 
radius.  If it does, then the effect of the stability criterion will also have
a radial dependence.  In practice, however, this is not of great
concern.  Numerous studies of late-type edge-on spirals in the optical
and infrared \citep[\eg][]{vdK82, sg90, dG96, dG97, fry99, bm02} have found 
that radial variations in the vertical scale heights of stars are always small,
$\lesssim 10\%$.  This is true even when the galaxies are tidally interacting:
a study by \citet{schwarz01} shows that the fractional variation of $z_0$ in 
disturbed edge-on systems is typically less than 14\%.  Thus, even if M33's 
disk is warped at large radii, our assumption of a constant value for
the scale height should be acceptable.

The solid lines of Figure~\ref{fig9} show the effect the stability
requirement has on our probability contours.  In the inner regions of the
galaxy, the requirement is satisfied almost everywhere:  virtually all the
solutions lie below the stability cutoff.  However, at large radii, the
criterion severely limits the allowable values for the vertical velocity
dispersion.  Such an effect is unavoidable:  if M33's disk is stable 
and isothermal, then the vertical velocity dispersion at large distances from 
the nucleus must be small.  

Figure~\ref{fig10} applies the stability criterion and marginalizes 
Figure~\ref{fig9}'s remaining probability contours over the parameters
$\sigma_R$ (the middle panel) and $\sigma_z$ (the lower panel).   The
top panel of the figure shows the behavior of $\sigma_z$ if one asserts
that M33's stellar disk is at the limit of stability; this solution
also represents the largest values of $\sigma_z$ possible under the 
isothermal disk approximation.  In the figure, we have illustrated the shapes
of the non-Gaussian errors via the widths of the shaded areas.  The figure
contains several features of note.

First, as anticipated, the requirement of stability produces a radial gradient 
in the vertical velocity dispersion.  Remarkably, this gradient is much 
shallower than that expected from a constant mass-to-light ratio disk.  
In the $V$-band, M33's disk scale length is $\sim 9\farcm 1$ \citep{guidoni}; 
in the infrared, where the effects of extinction gradients are less, the scale 
length shrinks to $\sim 5\farcm 8$ \citep{rv94}.  Our kinematic measurements 
imply a physical scale length for M33's disk that is $\sim 13\arcmin$ or 
$\sim 3.5$~kpc.  This scale length does not change much, even if one requires 
that the disk be pressing the limits of stability.

The other notable feature of Figure~\ref{fig10} is the behavior of the
radial component of M33's velocity ellipsoid.   Outside of $\sim 2.5$~kpc,
$\sigma_R$ declines with a scale length that is more than 3 times greater
than that of the galaxy's optical light; this is slightly larger than the
scale length one would expect from the survey data of \citet{bottema}, but
not exceptionally so.  However, inside of $\sim 2.5$~kpc, $\sigma_R$ 
levels off, and may even decline at small galactocentric radii.  Although
such behavior is not common, neither is it unique:  the line-of-sight velocity
dispersions of NGC~3198 and NGC~6503 suggest that these galaxies also have
low central values of $\sigma_R$ \citep{bottema}.

The turnover of M33's radial velocity dispersion is most likely a
manifestation of the requirement that $\sigma_R$ be much less than the 
galaxy's rotation speed.  M33's rotation velocity increases from zero near
the galactic center to $\sim 90$~\kms\ at $\sim 3$~kpc.  If $\sigma_R$
did not turn over, it would dominate galactic rotation within $\sim 1.5$~kpc 
of the nucleus, and the result would be a ``hot'' stellar population in the
galaxy's central regions.  If such a population exists in M33, it is
extremely weak \citep{bothun, rv94}.

The turnover of the radial velocity dispersion explains why the
system's line-of-sight dispersion is so flat:  the increase in $\sigma_z$
near the center of the galaxy is offset by the decline in $\sigma_R$.  This  
flattening, however, does not affect the large scale trend in the dispersion
ratio.  In the interior of the galaxy, $\sigma_z / \sigma_R \sim 0.6$; by 
$\sim 9$~kpc, the ratio has dropped to $\lesssim 0.4$.  This range of values
is consistent with the numerical models of disk heating computed by
\citet{villumsen} and \citet{jb90}.  More importantly, our inferred
$\sigma_z / \sigma_R$ gradient agrees with the analytic results of 
\citet{carlberg}, who showed that the velocity ellipsoid of a disk must
become rounder as the stellar velocity dispersion increases.  Our recovery
of this result supports the validity of our $\sigma_z$-$\sigma_R$ 
decomposition.

There is one more constraint that can be applied to M33's stellar
kinematics.  From the Jean's equation, the rotational velocity of stars
in a stellar disk is related to the circular velocity of the system's gas by
\begin{equation}
V_c^2 - \langle V_* \rangle^2 = \sigma_R^2 \left( {\sigma_{\phi}^2 \over 
\sigma_R^2} - 2 {\partial \ln \sigma_R \over \partial \ln R} - 
{\partial \ln \nu \over \partial \ln R} - 1 \right) -
R \, {\partial \sigma_{Rz} \over \partial z} - {R \over \nu} {\partial \nu
\over \partial z}
\end{equation}
where $\nu$ is the stellar density \citep{bt}.  If we use the 
epicyclic approximation, and assume that M33's PNe are scattered throughout
a constant scale height isothermal disk, then this equation for 
asymmetric drift simplifies to 
\begin{equation}
V_c^2 - \langle V_* \rangle^2 \simeq \sigma_R^2 \left( 
{1 \over 2} {\partial \ln V_c \over \partial \ln R} - 
2 {\partial \ln \sigma_R \over \partial \ln R} - 
2 {\partial \ln \sigma_z \over \partial \ln R} - {1 \over 2} \right)
\end{equation}
This equation, when combined with a model for the radial dependence of 
$\sigma_z$ and $\sigma_R$, places a further constraint on M33's galactic 
kinematics.

Based on the galaxy's luminosity profile and the data displayed in 
Figure~\ref{fig10}, it is reasonable to parameterize the radial dependence of
M33's vertical velocity dispersion with a simple exponential of scale 
length $R_z$.  An exponential function with scale length $R_R$ can also
be used to fit the galaxy's radial velocity dispersion, as long as the 
measurements are restricted to galactocentric distances greater than 
$\sim 2.5$~kpc.  Inside this radius, however, the exponential law breaks down, 
and the form of the $\sigma_R$ \vs\ $R$ relation is unknown.  

\begin{figure}
\figurenum{7}
\plotone{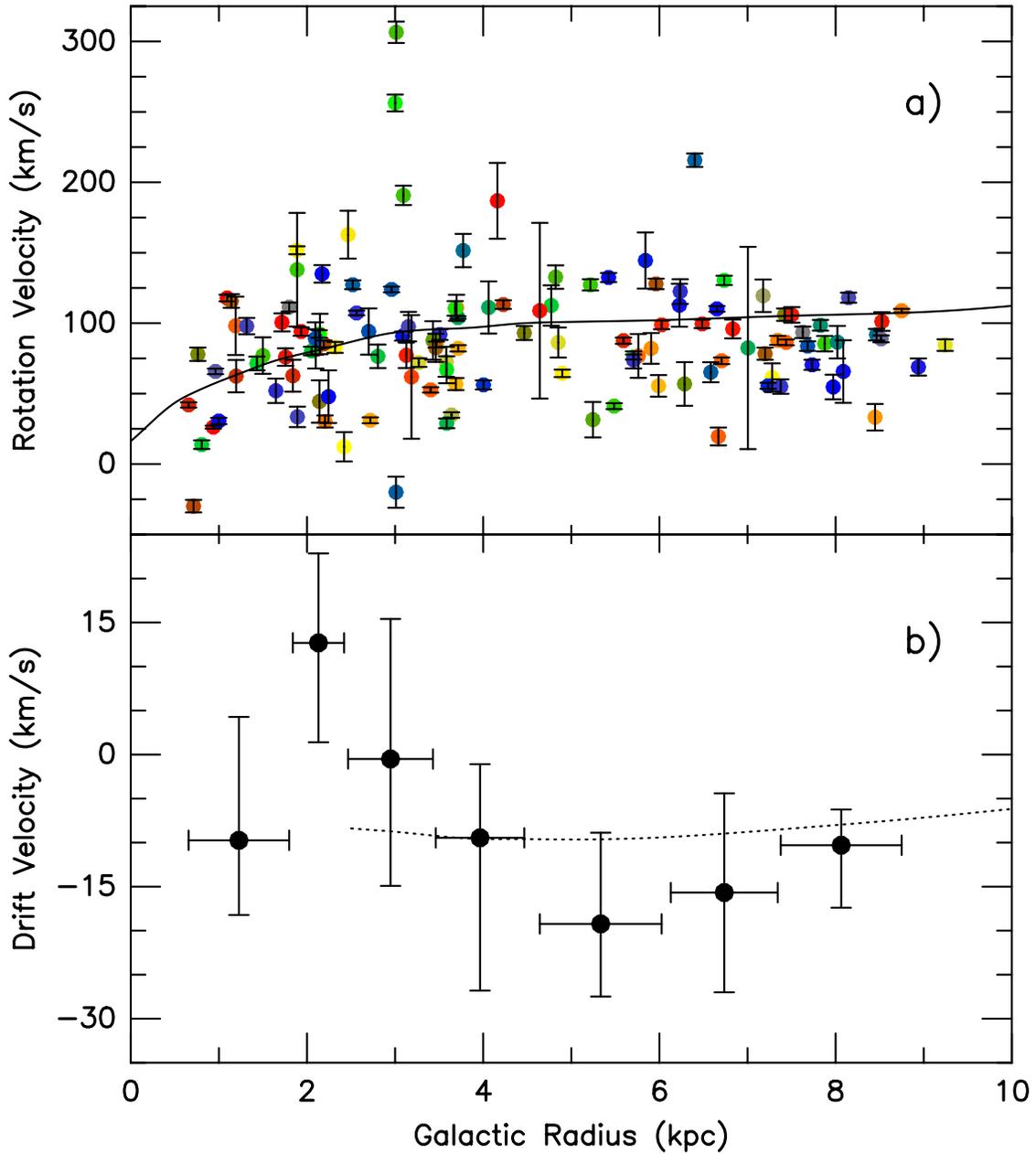}
\figcaption[fig7.eps]{The upper panel compares the inferred rotation speeds
of M31's planetaries with the rotation of the galaxy's H~I and CO gas.
The colors denote the azimuthal position of the PNe: the 
color sequence blue through green through red through yellow through blue
represents angles of $0^\circ$ through $90^\circ$ through $180^\circ$ through
$270^\circ$ through $360^\circ$ from the approaching major axis.   Fourteen
PNe projected within $10^\circ$ of the system's minor axis have 
not been plotted.   Note that the scatter in the diagram is significantly 
larger than the measurement errors for individual objects.  Note also that,
in the mean, the PNe rotate slightly slower than the gas; this is due to 
asymmetric drift.  The lower panel bins the data (with 18 PNe per bin)
to show the variation of the drift velocity with galactocentric radius.  Here, 
the error-bars in $y$ represent the $1\,\sigma$~uncertainty, as determined
via a bootstrap analysis, while the error bars in $x$ illustrate the size of 
the bin.  The dotted line is our best-fit model for regions outside 2.5~kpc 
(see Section~7). 
\label{fig7}}
\end{figure}

\begin{figure}
\figurenum{8}
\epsscale{0.9}
\plotone{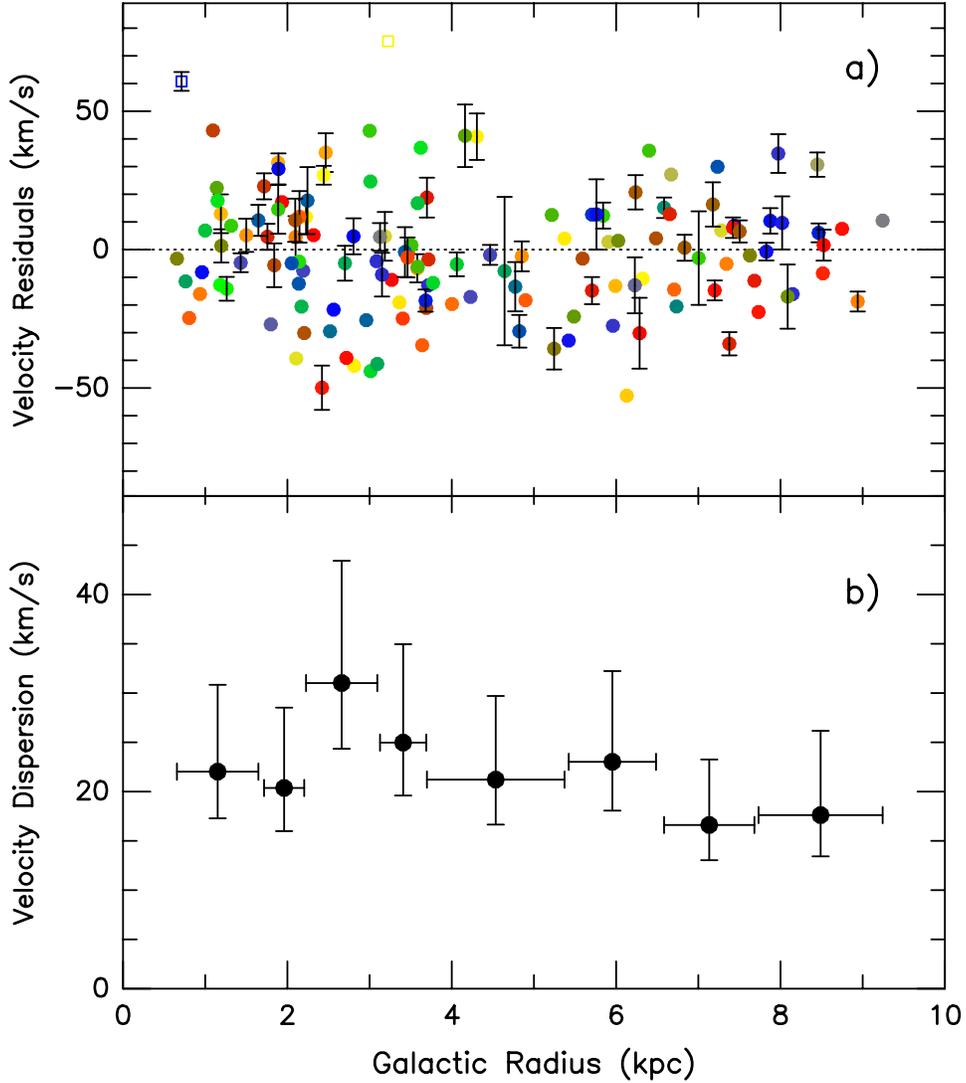}
\figcaption[fig8.eps]{The upper panel plots the PN radial velocities with the
galaxy's systemic velocity and disk rotation (defined as the H~I rotation 
speed minus asymmetric drift) removed.   Error bars have only been plotted for 
objects with random velocity uncertainties greater than 3~\kms; the two 
possible halo planetaries are shown as open squares.  As in Figure~\ref{fig7},
the colors represent the azimuthal position of each planetary, with the color
sequence blue through green through red through yellow through blue denoting
angles of $0^\circ$ through $360^\circ$ (referenced from the approaching major
axis).  The lower panel bins the data (with 18 PNe per bin) to show how the 
line-of-sight velocity dispersion varies with galactocentric radius.  The 
errors in $y$ represent 90\% confidence intervals, while the error bars in 
$x$ illustrate the size of each bin.
\label{fig8}}

\end{figure}

\begin{figure}
\figurenum{9}
\plotone{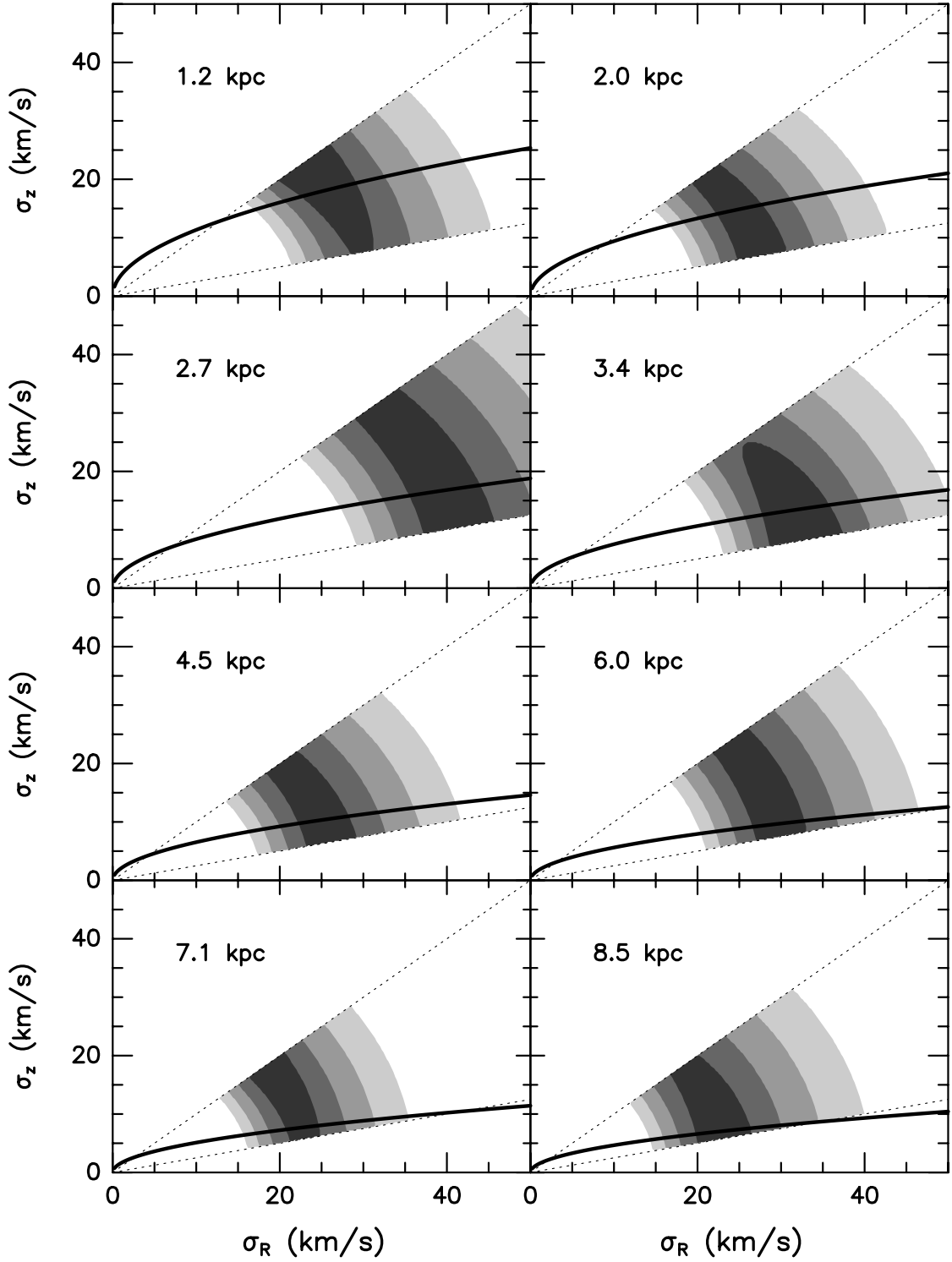}
\figcaption[fig9.eps]{Maximum likelihood solutions for the light-of-sight
radial velocity dispersions in the 8 radial bins displayed in Figure~8.
From dark to light, the shaded regions enclose 38\% ($0.5 \sigma$), 
68\% ($1 \sigma$), 86\% ($1.5 \sigma$), and 95\% ($2 \sigma$) of the 
probability.  The solid line displays the limits placed on the solutions 
by the requirement that the disk be stable against axisymmetric and 
non-axisymmetric perturbations; the dotted lines show the limits of our 
analysis.  Note that at large radii, the stability criterion strongly 
constrains the shape of the velocity ellipsoid and $\sigma_z$.
\label{fig9}}
\end{figure}

\begin{figure}
\figurenum{10}
\plotone{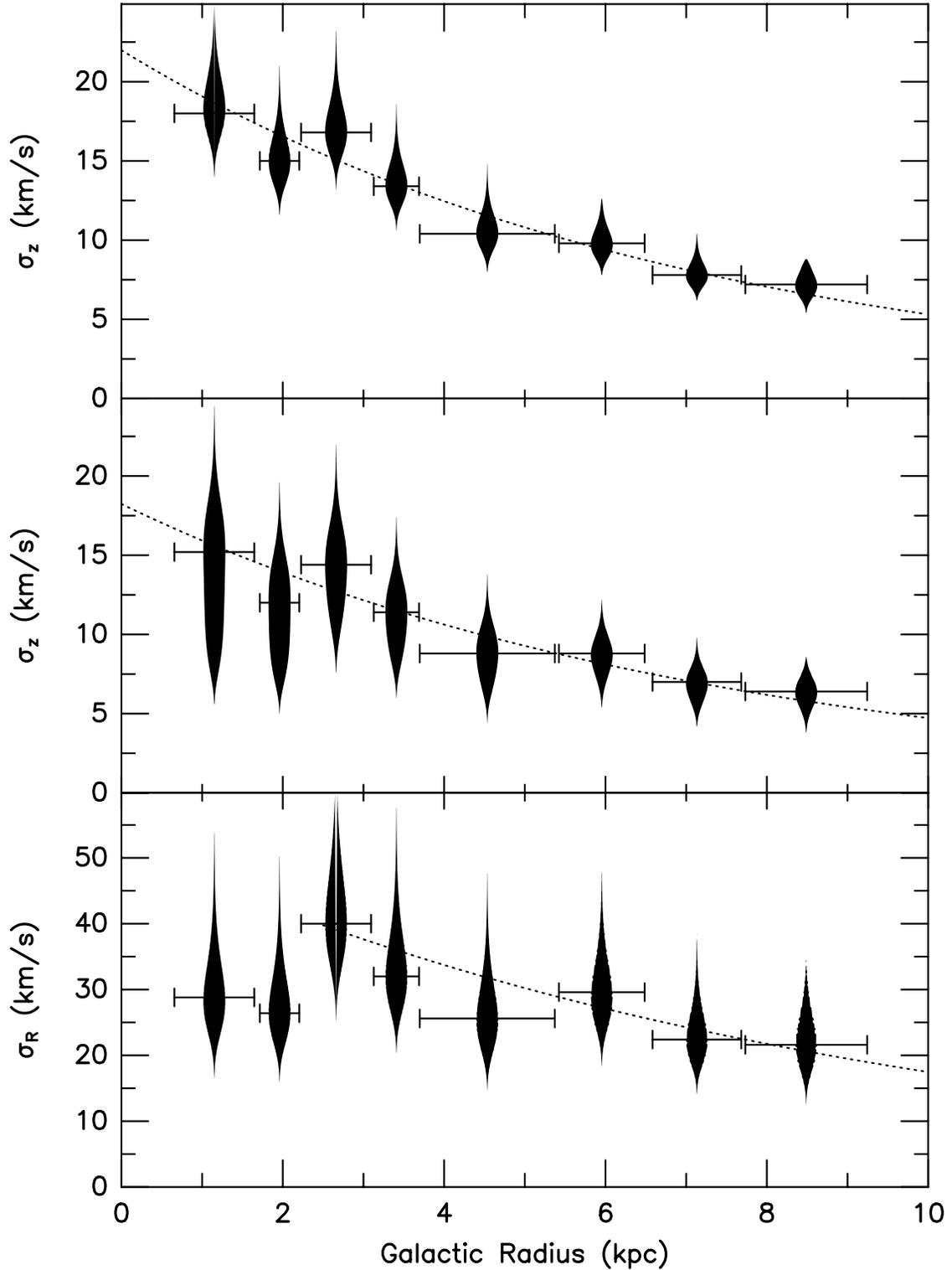}
\figcaption[fig10.eps]{The likelihoods of Figure~9, constrained by the
requirement of stability, and marginalized into one-dimensional plots in
$\sigma_z$ and $\sigma_R$.   The non-Gaussian uncertainties are illustrated
by the width of the lines.  The dotted lines show the best-fit exponentials,
when the additional constraints provided by our asymmetric drift measurements
are imposed (see text).   The middle and lower panels show total
likelihoods; the upper panel illustrates the solution for a marginally
stable (maximum mass) disk.  The implied scale length of $\sigma_z^2$ 
is more than twice as large as that of the $K$-band light.
\label{fig10}}
\end{figure}

Rather than guess the radial dependence of $\sigma_R$ in M33's inner regions,
we elected instead to exclude from the analysis all measurements of $\sigma_R$
and asymmetric drift within 2.5~kpc of the nucleus.  This allowed us to 
parameterize the radial dependence of both $\sigma_z$ and $\sigma_R$ with
exponentials and re-write (14) as 
\begin{equation}
V_c^2 - \langle V_* \rangle^2 \simeq \sigma_R^2 \left(
{2 R \over R_z} + {2 R \over R_R} - {1 \over 2} +
{1 \over 2} {\partial \ln V_c \over \partial \ln R} \right)
\end{equation}
Using this equation, we simultaneously solved for the values of $R_z$, $R_R$, 
$\sigma_R(R=0)$, and $\sigma_z(R=0)$ which best fit both the velocity 
dispersions of Figure~\ref{fig10} and the asymmetric drift data of 
Figure~\ref{fig7}.  To perform this fit, we minimized the $\chi^2$-like
statistic
\begin{equation}
\chi^2 = -2 \sum_i \ln \left\{ {P(y_i) \over P(y_m)} \right\} 
\end{equation}
where $P({y_i})$ is the probability of observing value $y_i$ from a 
probability distribution $P$, and $P(y_m)$ is the probability of observing the 
most-likely value.  This statistic, which reduces to the standard $\chi^2$ 
value for normally distributed errors, enabled us to consider the non-Gaussian 
uncertainties associated with $\sigma_R$ and $\sigma_z$, and the asymmetric
drift errors in a consistent manner.  It also permitted us to estimate the
uncertainties in our scale lengths via a modified jackknife analysis.  In this 
procedure, we repeatedly removed one planetary nebula at random from each of 
the bins, and re-fit the data.  When analyzed in this way, the best fit scale 
length for $\sigma_z^2$ became $13\farcm 5$ (with 90\% of the simulations lying
between $12\farcm 2$ and $14\farcm 9$), and the most likely value for the
scale length of $\sigma_R$ was $33\farcm 5$ (with 90\% of the simulations 
between $30\farcm 5$ and $38\farcm 6$).  The results do not change much if 
one forces the disk to marginal stability:  under this condition, the derived 
scale length of $\sigma_z^2$ declines by less than $\sim 5\%$.  Once again, 
these scale lengths are significantly larger than the $5\farcm 8$ value found 
from $K$-band photometry.  In fact, our $3 \,\sigma$ limits on M33's dynamical
scale length lie between $10\farcm 8$ and $16\farcm 0$, and none of our
20,000 simulations produced a scale length as low as the galaxy's optical
scale length of $9\farcm 1$.  Our best fit solution is illustrated in 
Figures~\ref{fig7}b and \ref{fig10} by dotted lines.

If the disk of M33 is indeed isothermal, then we can substitute the vertical
velocity dispersions of Figure~\ref{fig10} into equation (6) to produce 
estimates for the galaxy's disk mass surface density.  These densities can 
then be combined with optical \citep{guidoni} and infrared \citep{rv94} 
surface photometry to generate values for the total disk mass-to-light ratio. 
Finally, we can remove the contribution of M33's interstellar medium from the
mass-to-light values by subtracting the surface mass density of galactic H~I 
and H$_2$ gas \citep{heyer} from the total mass density, and applying a 
first-order correction for internal extinction, using the central extinction 
value ($A_V = 0.9$~mag) and extinction scale length ($\sim 10\farcm 2$) derived 
by \citet{rv94}.  These mass estimates and mass-to-light ratios are presented 
in Table~\ref{tab6}.

The errors quoted in Table~\ref{tab6} reflect only the formal uncertainties
associated with our measurement of $\sigma_z$; not included are the systematic 
errors associated with our adopted value of $z_0$ and the use of the 
isothermal disk approximation.  The former uncertainty is small: a 25\%
increase in the PN scale height (from 175~pc to 220~pc) results in only a
$\sim 5\%$ decrease in the derived values for M33's disk mass surface density
and scale length. 

The error associated with the isothermal disk approximation is slightly more
formidable.  Overall, the isothermal approximation provides an adequate fit to 
the vertical structure of edge-on spirals \citep{vdK82} and the stellar
kinematics of the solar neighborhood \citep[\eg][]{wielen}.   However,
near the galactic plane, the approximation breaks down:  the stellar
density distribution is more sharply peaked than the isothermal model would
predict \citep{fw87, dG96}.  This has led \citet{vdK88} to consider a family
of alternative models, which have isothermal (sech$^2 (z/z_0)$)
and exponential ($\exp (-z/z_e)$) spatial distributions as their limiting 
cases.   According to these models, M33's surface mass can be up to 33\%
larger than that derived using the simple isothermal assumption.   

The final systematic uncertainty that is not included in Table~\ref{tab6} is
one that affects our mass-to-light ratio measurements in M33's outer regions.  
Reliable optical and infrared surface photometry exist only for M33's inner 
$\sim 20\arcmin$; outside of this region, one must extrapolate the galaxy's 
surface brightness from its measured scale length.  The length of this 
extrapolation is significant, so a small error in scale length propagates
into a large error in $M/L$.  Moreover, if M33's disk is truncated 
\citep[\eg][]{kregel}, or if a small amount of excess luminosity is left over 
from a tidal interaction, then the derived mass-to-light ratios will be
seriously in error.  In fact, the extrapolated $V-K$ color for the outer
disk of M33 is not consistent of any population synthesis model
\citep[\eg][]{maraston}.  Given the uncertainties associated with infrared
photometry, our anomalously large values for $M/L_K$ in the outer disk of the
galaxy are almost certainly due to errors in the extrapolated $K$-band surface 
photometry.

\section{The Scale Length of M33's Disk}
Our value for M33's disk scale length is more than twice that of the galaxy's
infrared light, and 1.5~times that of the galaxy's $V$ light.
This discrepancy implies a gradient in the galaxy's disk
mass-to-light ratio.  Specifically, our observations imply that M33's disk
$M/L$ slowly increases with radius.  Part of this increase is due to the
growing contribution of M33's interstellar medium, which has a density
distribution that is much flatter than that of the stars \citep{heyer}.  
However, even when this component is taken into account, a positive gradient
the disk mass-to-light ratio remains.   The sign of this gradient runs
contrary to expectations: in most models of galaxy formation, it 
is the inner regions of a galaxy that form first, and contain the older, higher 
mass-to-light ratio stars \citep[\eg][]{els, sz, fe80, sm94}.  Moreover, 
multi-color imaging \citep{BdJ00} and H$\alpha$ surface photometry 
\citep{ryder} tend to support this picture: most disks appear to possess 
significant age gradients that are consistent with the inside-out scenario. 
Thus, we must consider the possibility that our mass measurements are biased. 
In fact, there are several effects which could conceivably introduce a 
systematic error into our estimates of $\sigma_z$.  If any of these exist, 
then our value for the galaxy's kinematic scale length could be in error.

The first possible bias comes from contamination by emission-line objects
that are not planetary nebulae.  In M33, H~II regions and supernova remnants
far outnumber planetary nebulae, and, as Population~I objects, their velocity 
dispersions are significantly less than that of the PNe \citep[\eg][]{zar89}. 
If the fraction of contaminants is higher near the center of M33 than it is
at large radii, then a systematic error in the scale length will be the
result.  Some support for this hypothesis comes from the image quality of the
[O~III] $\lambda 5007$ frame of M33's central field:  the seeing on that image
($1\farcs 05$) is the poorest of any frame in the \citet{survey} survey.

Still, for all the reasons detailed in Section~2, contamination is unlikely to 
be a significant problem in our survey.   All of our PN candidates were 
detected both in [O~III] $\lambda 5007$ and in H$\alpha$, and the excellent 
seeing on the latter image (and on the broadband $V$ frame) should have 
excluded virtually all H~II regions from our sample.  Similarly, given the 
survey's image quality, and the intrinsic ratio of PNe to supernovae, it is 
difficult to envision how the fraction of faint compact supernova remnants 
could be large enough to cause a systematic error in $R_z$.  At worst, 
contamination should cause only a slightly underestimate in $\sigma_z$ 
throughout the galaxy.

A second possible explanation for the large kinematic scale length involves
a selection effect.  The inner regions of M33 are quite dusty, with
internal extinction values of $A_V \sim 1$~mag \citep{israel, berkhuijsen, 
rv94, pg97}.  Consequently, PN surveys in this area are biased against objects 
in, and on the far side of the galactic plane.  Under the isothermal disk 
approximation, this bias is irrelevant, since the velocity dispersion of 
such a disk is independent of galactic latitude.  However, high above the 
plane, where the stellar orbits reach turn-around, the approximation must 
break down.   If most of the PNe in our sample are high latitude objects,
then it is possible that our PN velocities are not representative of the true
kinematic structure of the disk.

Since M33's internal extinction declines rapidly with
radius \citep{israel, berkhuijsen, rv94, pg97}, the above selection effect
has the potential for creating an artificial gradient in $\sigma_z$.  
However, the effect goes in the wrong direction.  If extinction is to
produce an overestimate of the disk's kinematic scale length, the PNe at 
high galactic latitude must have a lower dispersion that those near the
plane.  Yet models developed to fit the luminosity profiles of edge-on
galaxies all have vertical velocity dispersions that increase with galactic
latitude \citep{vdK88}.  So while internal extinction may indeed be
censoring our data, the effect cannot produce the observed discrepancy.

A third way of explaining our large value of $R_z$ involves our basic
assumptions about the kinematics of M33's disk and the stability criteria.
As Figure~\ref{fig9} demonstrates, the PN velocities, by themselves, 
do not fix $\sigma_z$.  Instead, it is the velocity dispersion measurements,
in combination with the requirements of dynamic stability, that produce
the galaxy's dispersion gradient.  If our warped-disk model for M33 is
incorrect, or if the stability criteria have an additional radial dependence,
or if the isothermal approximation breaks down at large radii, then an
error in the scale length could be the result.

The first of these possibilities can quickly be excluded:  in the area
covered by our survey, the effects of M33's warp are minimal.  In
fact, if we were to model the galaxy using a simple flat-disk geometry, 
the derived kinematic scale length of the system would {\it increase\/} by 
$\sim 10\%$.  

The latter two uncertainties are more difficult to assess.  Certainly, the
stability criterion given by (11) is only an approximation, as it assumes
spiral-arm pitch angles that are close to zero, and neglects (among other
things) the thickness of the disk and the existence of gaseous sub-systems. 
Nevertheless, numerical experiments suggest that the formulation
is approximately correct \citep[\eg][]{griv, khoperskov}.
Similarly, though the limit on $\sigma_z / \sigma_R$ from the buckling (or
fire-hose) instability is not precisely known, the numerical experiments to
date all indicate that a dispersion ratio of $\sim 0.3$ is reasonable
\citep{araki, ms94, sellwood}.  Finally, though one expects the isothermal
approximation to become less accurate at large radii (due to the longer
timescales for stellar interactions), surface photometry of edge-on spirals
suggests otherwise \citep[\eg][]{vdK82, fry99, bm02}.

The final way of creating an artificial gradient in M33's disk mass-to-light
ratio is through the introduction of a radial gradient in the PN scale height.
Our measurement of the disk scale length assumes that $z_0$ is constant
throughout the galaxy, and, for disks in general, this is a good assumption
\citep[\eg][]{sg90, dG96, dG97}.  However, our analysis also assumes that
PNe belong to the thin disk system and that the PN scale height does not
change with radius.

The former assumption is certainly valid.  Although deep observations of 
edge-on galaxies often reveal the existence of red ``thick disk'' 
envelopes \citep[\eg][]{dalcanton}, the amount of luminosity in this component
is small enough so that thick disk PNe should be exceedingly rare.  PN surveys
of the edge-on spirals NGC~891 \citep{p7} and NGC~4565 \citep{p10} confirm
this fact: very few objects are found in the thick disk region.  Still, within
the thin disk, the scale height of PNe could still change.  For example,
in the solar neighborhood, the vertical velocity dispersion of a stellar
population is roughly proportional to the square root of the population's
age \citep[\eg][]{wielen, jw83}.  Since planetary nebulae are derived from
a mix of populations with turnoff masses between $1 M_{\odot} \lesssim
M \lesssim 5 M_{\odot}$, their dispersion is a weighted average over the age
range which corresponds to these masses, \ie\ from $t_1 \sim 10^{10}$~yr to
$t_2 \sim 10^8$~yr.  In other words,
\begin{equation}
\langle \sigma_z^2 \rangle = \int_{t_1}^{t_2} \Phi(t) \, {d N \over
d m_{tn}} \, {d m_{tn} \over dt} \, \sigma_z^2(t) \,  dt 
\end{equation}
where $\Phi(t)$ is the region's star formation rate history, and $m_{tn}$ is
the main-sequence turnoff mass of a population with age $t$.  Since velocity
dispersion is related to scale height via equation (6), this means that a
gradient in a galaxy's star-formation rate history can produce an artificial
gradient in the PN scale height.

\begin{figure}
\figurenum{11}
\plotone{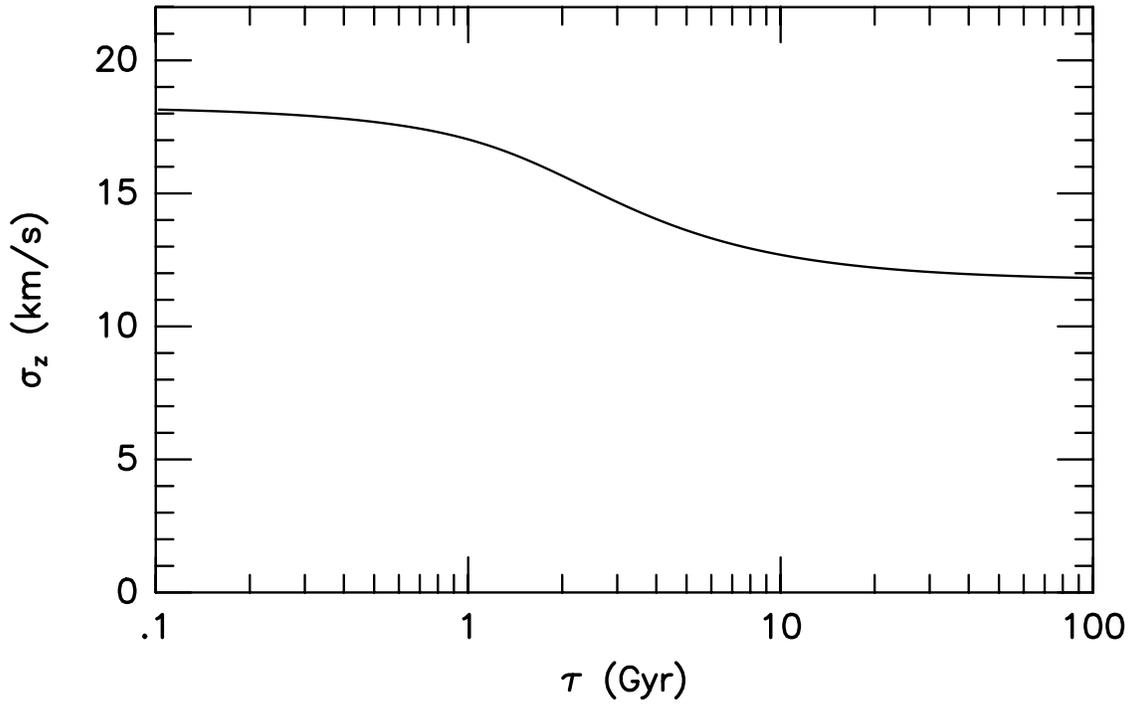}
\figcaption[fig11.eps]{The vertical velocity dispersion of disk planetary
nebulae as a function of the exponential decay time of star formation.
The plot assumes a \citet{scalo} initial mass function that is constant
with time, the \citet{iben} initial mass-stellar lifetime relation, and that
$\sigma_z \propto t^{1/2}$.  The $y$-axis has been arbitrarily normalized 
to 20~\kms\ at $10^{10}$~yr.  Note that disks with a constant star formation 
rate ($\tau \gg 1$) have velocity dispersions that are a third smaller than 
disks whose star formation has long ago decayed away.
\label{fig11}}
\end{figure}

Figure~\ref{fig11} quantifies this effect for the case where $\Phi$ declines
exponentially with scale time $\tau$.  In late-type spirals such as M33, the 
evidence suggests that the rate of star formation has been roughly constant 
over a Hubble time \citep{ktc}.  If this is the case, then $\tau \gg 1$, and 
the dependence of $\sigma_z$ on $\Phi$ is weak.  Nevertheless, if the
star-formation rate history of M33's inner disk is substantially different
from that of its outer disk, a systematic error in our mass determinations
may be the result.  

Unfortunately, the effect shown in Figure~\ref{fig11} is much too small to
explain M33's large kinematic scale length:  in order to be consistent with a
constant $M/L_V$ disk, the scale height of M33's PNe would have to change by
almost an order of magnitude, from $\gtrsim 500$~pc near the galactic center,
to $\lesssim 100$~pc at 8~kpc.  Not only is this many times greater than
what can be achieved by changing the galaxy's star formation rate history,
but the direction of the effect is wrong.  If the stars of M33's interior are
systematically older than the stars of the outer disk (as might be expected
from an inside-out galaxy formation scenario), then our assumption of a 
constant $z_0$ would produce an overestimate of the surface mass in the
inner regions of the galaxy, and an underestimate of surface mass in the outer
disk.  This would cause us to underestimate the scale length of the system.
Moreover, the discrepancy between the galaxy's photometric and dynamical scale 
lengths would be exacerbated by the reaction of the system's infrared
light to such a gradient.  While the $K$-band is far superior to the optical
when in comes to viewing stellar luminosity through the veil of interstellar
extinction, it is not immune from population effects.  Just as the OB stars
of present day star formation distort surface brightness measurements made
in $B$, red supergiants left over from star-formation of the recent past
contaminate data taken in $K$.  Although photometric enhancements in the 
infrared are less than those in the optical, they are still substantial 
\citep{rv94}, and if the interior regions of M33 are older than the stars
of the outer disk, the galaxy's IR luminosity profile will appear more 
diffuse that the mass.  This is the opposite of what is seen.

We note that, in principle, the question of M33's star formation history
can be answered via planetary nebula spectrophotometry.  \citet{kj90} have 
shown that the ratio of nitrogen to oxygen in a planetary nebula correlates 
with its core mass:  objects with $M_c < 0.65 M_{\odot}$ have N/O $\sim 0.2$, 
while higher mass objects have $1 \lesssim {\rm N/O} \lesssim 2$.  
Consequently, this discontinuity, which is most likely caused by dredge-up in 
AGB stars with $M \gtrsim 3 M_{\odot}$ \citep{ir83}, can be used to identify 
the PNe of high mass progenitors.  Furthermore, \citet{dopita} have shown 
that one can estimate PN central star masses directly by using deep 
spectrophotometry to locate the stars on the HR diagram, and then comparing
their locations to post-AGB evolutionary models.  This technique, when
combined with an initial-mass final-mass relation \citep[\ie][]{weidemann},
can allow us to trace the star-formation history of different parts of M33's
disk, and test for the presence of age gradients in the PN sample.  

\section{M33's Disk and Halo}
There have been numerous efforts to probe the radial distribution of M33's
dark halo via the shape of the galaxy's rotation curve \citep[\eg][]{g00, 
cs00, c03}.  However, these efforts have all been limited by the
unknown mass-to-light ratio of the galaxy's disk.   The disk-mass estimates
of Table~\ref{tab6} provide a new constraint on M33's mass distribution, and
help break the disk-halo degeneracy.  

\begin{figure}
\figurenum{12}
\plotone{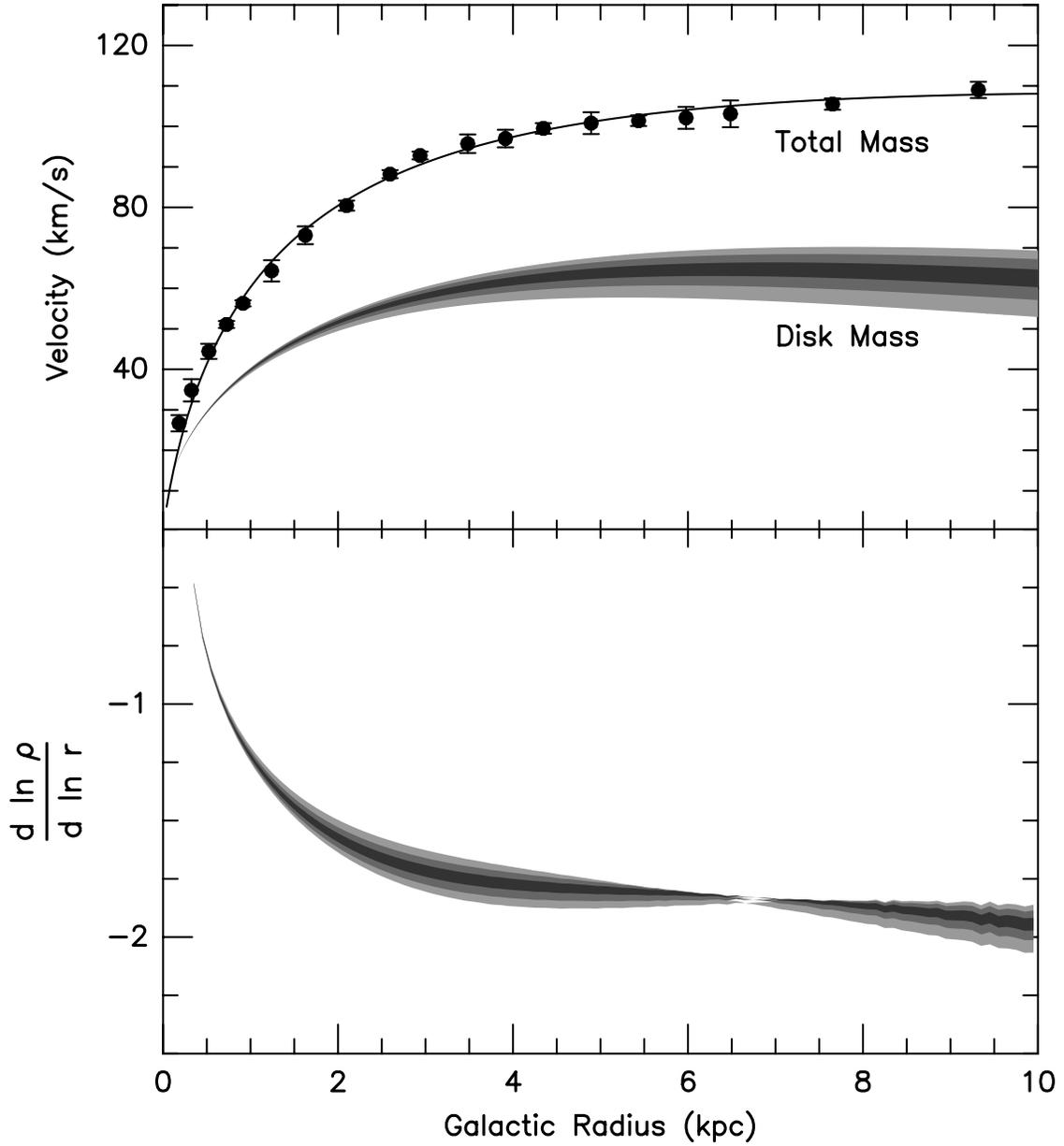}
\figcaption[fig12.eps]{The upper panel compares M33's H~I and CO rotation
curve to the rotation curve which would be produced from the disk alone.  
The solid line is a \citet{brandt} law fit to the H~I and CO velocity
measurements; the shaded regions indicated the 1, 2, and $3 \, \sigma$
uncertainties about our best-fit mass model (as determined from a
jackknife analysis).  The lower panel shows the density law inferred for 
the galaxy's dark halo (under the assumption of spherical symmetry), again
with 1, 2, and $3 \, \sigma$ uncertainties.
\label{fig12}}
\end{figure}

To demonstrate this constraint, we began by fitting a smooth \citet{brandt} 
curve to the H~I and CO rotation measurements of \citet{c03}.  We then used 
our mass model to infer a rotation curve of the disk alone.  The radial profile
of the dark halo was then estimated by simply differencing the two components.
Figure~\ref{fig12} illustrates this disk-halo decomposition. 

From the rotation curves of Figure~\ref{fig12} and the mass-to-light ratios
of Table~\ref{tab6}, it is clear that there is very little dark matter in
M33's central regions.  Based on our PN velocity dispersion measurements,
the disk mass-to-light ratios for M33's central $\sim 2$~kpc are
$M/L_K \sim 0.2$ and $M/L_V \sim 0.3$.  Both these values are close to those
predicted for stellar populations dominated by young ($\tau \sim 150$~Myr 
old) stars \citep{maraston, BdJ01}.  More importantly, these numbers are 
statistically identical to those obtained using the galaxy's H~I and CO 
rotation curve, and are also similar to the central mass-to-light ratios 
derived from the rotation curves of other spirals \citep[\eg][]{moriondo}. 
This consistency implies that the constant mass-to-light approach of 
``maximal disk'' models is tenable, at least in the inner regions of galaxies.

Outside of $\sim 2$~kpc, M33's disk mass-to-light ratio exhibits a small, but
distinct gradient, going from $M/L_V \sim 0.3$ at $\sim 2$~kpc to $M/L_V 
\sim 1.5$ at $\sim 8.5$~kpc.  Since this appears to be in conflict with the
inside-out model of galaxy formation, one possible explanation is the
presence of dark matter associated with M33's disk.  Indeed, M33's disk-mass
alone produces a rotation curve that is more-or-less flat with radius.  

Another explanation is the increasing importance of M33's gaseous disk.
The surface density of M33's gas declines more slowly than that of galaxy's
optical light \citep{rv94, c03}.  Consequently, at large radii, this
component can be expected to dominate the disk's mass budget.  However, as
summarized in Table~6, the radius where this occurs lies significantly outside
the region of our survey.  Even in our outermost bin, M33's interstellar
medium comprises only $\sim 25\%$ of the disk mass.

One final explanation for the behavior of M33's disk mass-to-light ratio lies
in the vigorous star formation occurring in M33's inner regions. It is
entirely possible that, near the center of the galaxy, the mass contributed by 
the region's older, higher mass-to-light ratio populations is being overwhelmed
by the luminosity of present-day stars.  Evidence for this possibility comes 
from the results of \citet{c03} and \citet{heyer}, who demonstrate that
star formation rate of M33's disk has a steeper radial dependence than
the galaxy's $V$-light.

Whatever the reason for M33's increasing disk mass-to-light ratio, the fact 
that the galaxy's disk mass-to-light ratios are consistent with those expected 
from young stellar populations strongly suggests that dark matter is not an 
important constituent of the disk.  Models, such as the molecular gas 
hypothesis of \citet{pfenniger}, are certainly ruled out.

Our analysis is based on the velocities of only 140~PNe, so the spatial 
resolution of our measurements is poor.  However, if we assume that M33's 
disk has an exponential profile, and that our surface mass densities 
are accurate, then we can use our disk/halo decomposition to place limits
on the radial profile of the galaxy's dark matter.  According to Cold Dark
Matter models of galaxy formation, systems such as M33 should contain
``cusps'' of dark matter at their center, with power-law density profiles
in the range $-1 > \alpha > -1.5$, where $\rho_h \propto r^\alpha$
\citep[\eg][]{nfw, moore, dave, klypin, power}.  As is illustrated in
the lower panel of Figure~\ref{fig12}, our central slope is much shallower
than this, with $\alpha \gtrsim -0.5$.  This result is robust:  if we alter
the PN scale height by 50\%, or change the disk scale length by 25\%, 
or increase the total disk mass 33\% (either by adopting an exponential 
law for the disk's vertical profile, or by changing the central value of 
$\sigma_z$), the slope of M33's dark matter core remains low.  The result
also agrees with the conclusions of \citet{c03}, who ruled out the presence 
of a steep dark matter core in M33 on the basis of the galaxy's rotation
curve, and the assumption of a constant ($K$-band) mass-to-light ratio disk.
Our analysis removes the latter assumption, but is otherwise similar.

If our analysis is correct, then M33 is another example of a galaxy whose 
interior dark matter profile is flat.  High resolution rotation curves
now exist for a large number of low-luminosity and low-surface 
brightness spirals and dwarfs, and every one is consistent with a value of 
$\alpha > -1$ \citep[\eg][]{bs01, swaters}.  It is still unclear whether 
this shallow a slope is a critical problem for the Cold Dark Matter paradigm 
\citep{dBBMc}, or whether the discrepancy can be explained by the limitations
in the measurements or the models \citep{swaters, spekkens}.   Nevertheless, 
M33 appears to be another example this problem.

Outside of $\sim 1$~kpc, our analysis suggests that M33's dark matter halo 
obeys the inverse-square law density dependence that is expected from 
isothermal and \citet{nfw} distributions.  Because the dynamical scale length
of the disk is larger than the photometric scale length, the system's
total disk mass is larger than that expected from the maximal disk
hypothesis.  Within a radius of 10~kpc, our velocity dispersion measurements 
yield a disk mass $\sim 9.1 \times 10^{9} M_{\odot}$.  This is 
$\sim 2.5$~times larger than one would obtain using $V$-band surface
photometry and the maximum disk hypothesis, and $\sim 6$~times larger than
one would derive from a similar analysis in $K${}.  Still, within this
radius, M33's disk contains only $\sim 30\%$ of the total mass of the galaxy.

Of course, M33's disk does not end at 10~kpc.  Although our planetary
nebula detections end near this 10~kpc limit, observations of M33's H~I
gas extend to nearly twice that distance \citep{c03}.   The interpretation of 
the extended H~I data is made slightly more difficult by the increasing
importance of the galactic warp at large radii.  Nevertheless, if the models of
\citet{cs97} and \citet{cs00} are accurate, then M33's rotational velocity
increases from $\sim 110$~\kms\ to  $\sim 135$~\kms\ between 10~kpc and 20~kpc. 
This suggests that the bulk of M33's dark matter halo still resides outside 
the region of our survey.

\section{Conclusions}
Until now, our knowledge of the stellar kinematics of spiral galaxies has
come principally from observations of stars in the solar neighborhood, and
secondarily from absorption-line measurements in the inner regions of
face-on and edge-on spirals.  Our planetary nebula observations have
improved upon this situation.  For the first time, we have been able to
measure the shape of the stellar velocity ellipsoid throughout a galactic
disk, and directly measure the variation in a galaxy's disk mass surface
density.  Our data suggest that, in the inner regions of galaxies, maximal 
disk models for the rotation curve are appropriate, and the contribution of
dark matter is negligible.  However, the data also show that the 
mass-to-light ratios of galactic disks do change with radius, and the $M/L$ 
of the inner disk may be a factor of $\gtrsim 4$ smaller than that of the 
outer disk.  Still, the existence of this gradient does not change the 
fact that most of a galaxy's material lies well away from the disk in a
dark halo that roughly follows an inverse square density law.  Models in
which the dark matter is in the form of molecular gas are excluded by our 
observations.

\acknowledgments
We would like to thank E. Corbelli for providing M33's rotation curve in
digital form, D. Willmarth for assisting in the setup for our PN
spectroscopy, and  R. Chandar for providing additional PN spectra for our
analysis.  In addition, we would like to the thank M. Bershady, J. Sellwood, 
S. Sigurdsson, and S. McGaugh for useful comments during the preparation of 
this paper.  This work was supported by NSF grant AST 00-71238.

\clearpage
\begin{deluxetable}{lccccccc}
\tablewidth{0pt}
\tablecaption{M33 Survey}
\tablehead{
& & & & &\colhead{Number of} &\colhead{Total Exp} & \\
\colhead{Field} &\colhead{$\alpha(2000)$} &\colhead{$\delta(2000)$}
&\colhead{Filter} &\colhead{Bandpass}
&\colhead{Exposures} &\colhead{Time (min)} 
&\colhead{Seeing} \\ }
\startdata
M33 Center &1:33:49.0 &+30:40:00 &[O~III]   &55~\AA  &5 &25 &$1\farcs 05$ \\
M33 Center &1:33:49.0 &+30:40:00 &H$\alpha$ &80~\AA  &5 &25 &$0\farcs 90$ \\
M33 Center &1:33:49.0 &+30:40:00 &$V$       &940~\AA &5 &1 &$0\farcs 95$ \\
M33 Center &1:33:49.0 &+30:40:00 &$B$       &990~\AA &5 &1 &$0\farcs 95$ \\ 
\\
M33 North &1:34:00.0 &+30:55:37 &[O~III]    &55~\AA  &5 &25 &$0\farcs 90$ \\
M33 North &1:34:00.0 &+30:55:37 &H$\alpha$  &80~\AA  &5 &25 &$0\farcs 70$ \\
M33 North &1:34:00.0 &+30:55:37 &$V$        &940~\AA &5 &1  &$1\farcs 05$ \\
M33 North &1:34:00.0 &+30:55:37 &$B$        &990~\AA &5 &1  &$1\farcs 05$ \\ 
\\
M33 South &1:33:10.5 &+30:22:30 &[O~III]    &55~\AA  &5 &25 &$0\farcs 95$ \\
M33 South &1:33:10.5 &+30:22:30 &H$\alpha$  &80~\AA  &5 &25 &$0\farcs 75$ \\
M33 South &1:33:10.5 &+30:22:30 &$V$        &940~\AA &5 &1  &$0\farcs 85$ \\
M33 South &1:33:10.5 &+30:22:30 &$B$        &990~\AA &5 &1  &$0\farcs 85$ \\ 
\enddata
\label{tab1}
\end{deluxetable}

\clearpage
\begin{deluxetable}{rcccccccccccl}
\tablewidth{0pt} \tabletypesize{\scriptsize} 
\tablecaption{M33 Planetary Nebula Candidates} 
\tablehead{ 
\colhead{ID} 
&\colhead{$\alpha$(2000)} 
&\colhead{$\delta$(2000)} 
&\colhead{m$_{5007}$} 
&\colhead{n$_{\rm p}$\tablenotemark{a}}
&\colhead{log($F_{H\alpha}$)\tablenotemark{b}} 
&\colhead{$\frac{F({\rm [O~III]})}{F({{\rm H}\alpha})}$ } 
&\colhead{err} 
&\colhead{n$_{\rm s}$\tablenotemark{c}} 
&\colhead{$v_{\odot}$}  
&\colhead {$\sigma_v$} 
&\colhead{M01\tablenotemark{d}}
&\colhead{Notes}
}
\startdata
   1 & 1:32:09.04 & 30:22:05.7 & 23.51 & 1 & $-15.321$ & 2.628 & 0.176 & 2 &$-139.2$ & 3.6&&\\
   2 & 1:32:26.54 & 30:25:49.8 & 23.53 & 2 & $-15.390$ & 3.042 & 0.204 & 2 &$-127.7$ & 1.9  & 1 &O\\
   3 & 1:32:38.03 & 30:24:00.6 & 21.82 & 2 & $-14.243$ & 1.046 & 0.019 & 2 &$-130.3$ & 1.6  & 2&O\\
   4 & 1:32:39.80 & 30:37:41.0 & 22.07 & 2 & $-14.602$ & 1.891 & 0.047 & 3 &$-208.4$ & 1.7  & 3&\\
   5 & 1:32:40.08 & 30:35:15.3 & 23.38 & 2 & $-15.343$ & 3.125 & 0.154 & 2 &$-160.4$ & 2.7  &&\\
   6 & 1:32:42.72 & 30:12:25.5 & 22.63 & 1 & $-14.895$ & 2.218 & 0.085 & 2 &$-106.7$ & 2.0  & 4&\\
   7 & 1:32:44.29 & 30:43:25.9 & 24.74 & 2 & $-15.655$ & 1.837 & 0.153 & 3 &$-185.9$ & 3.0  & 6&\\
   8 & 1:32:49.16 & 30:51:09.1 & 21.45 & 2 & $-14.526$ & 2.826 & 0.042 & 3 &$-192.2$ & 2.0  & 7&\\
   9 & 1:32:52.96 & 30:41:57.7 & 22.49 & 2 & $-14.778$ & 1.930 & 0.041 & 2 &$-167.3$ & 3.0  & 9&\\
  10 & 1:32:53.72 & 30:32:25.8 & 23.73 & 2 & $-14.720$ & 0.541 & 0.023 & 3 &$-136.3$ & 5.4 & 12&O\\
  11 & 1:32:55.04 & 30:09:52.9 & 21.34 & 1 & $-14.437$ & 2.549 & 0.048 & 2 &$-88.5$ & 1.1 & 13&\\
  12 & 1:32:55.11 & 30:14:01.5 & 22.23 & 1 & $-14.859$ & 2.948 & 0.090 & 2 &$-121.1$ & 3.0&&\\
  13 & 1:32:58.80 & 30:27:38.0 & 22.28 & 2 & $-14.887$ & 3.009 & 0.089 & 2 &$-136.0$ & 1.8 & 14&\\
  14 & 1:32:59.40 & 30:10:24.0 & 22.84 & 1 & $-14.459$ & 0.671 & 0.019 & 2 &$-94.5$ & 5.6 & 15&\\
  15 & 1:33:01.25 & 30:15:31.1 & 23.42 & 1 & $-15.210$ & 2.212 & 0.171 & 2 &$-114.7$ & 3.6 & 16&O\\
  16 & 1:33:05.62 & 30:42:15.1 & 23.43 & 2 & $-15.323$ & 2.863 & 0.132 & 0 &$-136.6$ & 8.4 &&O\\
  17 & 1:33:05.86 & 30:50:44.2 & 23.48 & 2 & $-15.574$ & 4.848 & 0.239 & 3 &$-201.8$ & 2.9&&\\
  18 & 1:33:06.11 & 30:31:04.5 & 20.72 & 2 & $-14.289$ & 3.203 & 0.050 & 3 &$-145.1$ & 2.0 & 18&O\\
  19 & 1:33:06.20 & 31:00:56.1 & 24.14 & 1 & $-15.823$ & 4.669 & 0.513 & 4 &$-195.5$ & 4.4&&\\
  20 & 1:33:07.44 & 30:54:23.2 & 21.14 & 2 & $-14.396$ & 2.769 & 0.038 & 5 &$-187.1$ & 2.3 & 17&\\
  21 & 1:33:08.55 & 30:34:38.5 & 23.96 & 2 & $-15.705$ & 4.211 & 0.526 & 2 &$-145.1$ & 5.1&&\\
  22 & 1:33:08.99 & 30:13:58.6 & 23.78 & 1 & $-15.304$ & 1.973 & 0.205 & 2 &$-91.1$ & 3.7 & 19&O\\
  23 & 1:33:11.96 & 30:30:48.2 & 22.74 & 2 & $-14.954$ & 2.301 & 0.095 & 3 &$-156.8$ & 1.3&&\\
  24 & 1:33:14.12 & 30:40:42.7 & 21.68 & 2 & $-14.630$ & 2.897 & 0.065 & 5 &$-188.3$ & 2.9 & 22&\\
  25 & 1:33:15.50 & 30:17:53.7 & 24.51 & 1 & $-15.594$ & 1.972 & 0.236 & 1 &$-133.0$ & 12.8 & 25&\\
  26 & 1:33:16.89 & 30:29:59.9 & 22.81 & 2 & $-14.928$ & 2.025 & 0.060 & 3 &$-119.0$ & 1.5 & 26&\\
  27 & 1:33:17.01 & 30:29:59.8 & 24.53 & 1 & $-15.848$ & 3.464 & 0.536 & 1 &$-120.6$ & 7.2 & 29&\\
  28 & 1:33:18.48 & 30:12:38.8 & 23.23 & 1 & $-15.276$ & 3.076 & 0.174 & 2 &$-110.6$ & 2.8 & 27&\\
  29 & 1:33:19.25 & 30:29:40.2 & 20.93 & 2 & $-14.417$ & 3.522 & 0.057 & 2 &$-141.1$ & 1.4 & 28&\\
  30 & 1:33:19.73 & 30:43:05.5 & 22.19 & 2 & $-14.476$ & 1.273 & 0.023 & 3 &$-108.7$ & 1.5 & 35&H, O\\
  31 & 1:33:21.16 & 31:06:44.0 & 22.83 & 1 & $-15.104$ & 3.006 & 0.113 & 2 &$-231.1$ & 2.5 & 30&O\\
  32 & 1:33:21.18 & 30:41:14.6 & 21.78 & 2 & $-14.534$ & 2.116 & 0.051 & 2 &$-216.5$ & 1.4 & 31&\\
  33 & 1:33:21.59 & 30:36:57.2 & 23.22 & 2 & $-14.760$ & 0.944 & 0.032 & 2 &$-112.8$ & 7.0 & 32&\\
  34 & 1:33:22.85 & 30:13:41.0 & 22.59 & 1 & $-14.791$ & 1.823 & 0.067 & 2 &$-134.9$ & 4.2 & 33&\\
  35 & 1:33:23.65 & 30:19:39.8 & 23.59 & 1 & $-15.419$ & 3.076 & 0.223 & 3 &$-120.1$ & 4.9&&\\
  36 & 1:33:24.23 & 30:16:14.5 & 23.65 & 1 & $-15.214$ & 1.813 & 0.141 & 3 &$-90.2$ & 1.6 & 36&\\
  37 & 1:33:25.11 & 30:45:06.2 & 24.45 & 2 & $-14.523$ & 0.176 & 0.013 & 4 &$-192.4$ & 8.8 & 130&\\
  38 & 1:33:26.33 & 30:41:45.2 & 22.36 & 2 & $-14.490$ & 1.118 & 0.024 & 2 &$-153.3$ & 3.4 & 41&\\
  39 & 1:33:26.59 & 30:35:50.3 & 22.41 & 2 & $-14.308$ & 0.705 & 0.019 & 4 &$-135.4$ & 7.6 & 38&\\
  40 & 1:33:27.29 & 30:40:51.3 & 22.79 & 2 & $-14.390$ & 0.598 & 0.014 & 2 &$-163.1$ & 6.2 & 37&O\\
  41 & 1:33:27.79 & 30:34:29.2 & 22.65 & 2 & $-14.442$ & 0.771 & 0.020 & 3 &$-120.2$ & 9.3 & 40&\\
  42 & 1:33:27.89 & 31:06:22.3 & 25.45 & 1 & $-15.899$ & 1.673 & 0.236 & 3 &\nodata & &&O\\
  43 & 1:33:28.50 & 30:37:45.8 & 22.24 & 2 & $-14.901$ & 3.239 & 0.098 & 2 &$-120.8$ & 1.1 & 39&\\
  44 & 1:33:31.25 & 30:40:04.9 & 24.11 & 2 & $-15.767$ & 4.238 & 0.698 & 0 &  & &&NT\\
  45 & 1:33:31.88 & 30:30:32.6 & 22.53 & 2 & $-14.604$ & 1.245 & 0.035 & 3 &$-155.9$ & 1.6 & 45&\\
  46 & 1:33:32.17 & 30:28:20.2 & 20.95 & 2 & $-14.247$ & 2.344 & 0.030 & 2 &$-123.8$ & 2.4 & 42&\\
  47 & 1:33:32.21 & 30:31:46.1 & 23.22 & 2 & $-14.564$ & 0.603 & 0.032 & 2 &$-170.6$ & 8.0&&\\
  48 & 1:33:32.34 & 30:42:40.4 & 21.89 & 2 & $-14.607$ & 2.264 & 0.063 & 2 &$-229.4$ & 2.4 & 49&\\
  49 & 1:33:32.69 & 30:26:42.3 & 22.53 & 2 & $-14.409$ & 0.794 & 0.019 & 3 &$-93.7$ & 7.2 & 43&\\
  50 & 1:33:33.15 & 30:37:34.4 & 22.15 & 2 & $-14.235$ & 0.759 & 0.015 & 3 &$-144.3$ & 6.0&&\\
  51 & 1:33:36.79 & 30:31:40.5 & 22.91 & 2 & $-15.218$ & 3.611 & 0.205 & 2 &$-115.9$ & 3.0 & 48&\\
  52 & 1:33:36.83 & 30:38:45.6 & 23.68 & 3 & $-15.160$ & 1.549 & 0.096 & 1 &$-146.9$ & 7.0&&\\
  53 & 1:33:37.00 & 30:26:33.1 & 21.19 & 2 & $-14.309$ & 2.170 & 0.031 & 2 &$-117.4$ & 1.7 & 46&\\
  54 & 1:33:38.56 & 30:33:02.3 & 22.50 & 2 & $-14.736$ & 1.738 & 0.047 & 3 &$-107.5$ & 2.3 & 51&\\
  55 & 1:33:40.16 & 30:37:49.5 & 21.81 & 3 & $-14.572$ & 2.249 & 0.040 & 1 &$-165.2$ & 0.6&&\\
  56 & 1:33:42.30 & 30:37:39.7 & 21.70 & 2 & $-14.433$ & 1.813 & 0.049 & 2 &$-171.0$ & 2.0 & 54&\\
  57 & 1:33:42.76 & 30:49:05.6 & 23.21 & 2 & $-14.802$ & 1.053 & 0.035 & 5 &$-222.7$ & 5.0 & 58&O\\
  58 & 1:33:43.48 & 30:59:41.1 & 24.42 & 1 & $-14.583$ & 0.209 & 0.012 & 3 &$-254.6$ & 10.1 & 57&\\
  59 & 1:33:43.75 & 30:33:26.6 & 22.67 & 2 & $-14.920$ & 2.257 & 0.085 & 2 &$-122.7$ & 4.7&&\\
  60 & 1:33:44.38 & 30:20:23.7 & 21.15 & 1 & $-14.409$ & 2.844 & 0.060 & 2 &$-115.7$ & 1.6 & 59&\\
  61 & 1:33:44.62 & 31:04:03.5 & 25.21 & 1 & $-15.886$ & 2.023 & 0.225 & 2 &\nodata  & &&O\\
  62 & 1:33:45.12 & 31:08:33.5 & 25.54 & 1 & $-16.032$ & 2.085 & 0.342 & 3 &\nodata  & &&O\\
  63 & 1:33:45.62 & 30:25:56.3 & 23.54 & 2 & $-15.389$ & 3.001 & 0.179 & 0 &\nodata  & &&O\\
  64 & 1:33:46.09 & 30:33:35.8 & 23.11 & 2 & $-14.854$ & 1.307 & 0.074 & 1 &$-106.4$ & 4.7&&\\
  65 & 1:33:46.45 & 30:26:55.4 & 22.40 & 2 & $-14.884$ & 2.682 & 0.079 & 3 &$-140.1$ & 3.0 & 60&\\
  66 & 1:33:46.73 & 30:17:33.5 & 21.15 & 1 & $-14.426$ & 2.940 & 0.069 & 2 &$-107.1$ & 2.2 & 61&\\
  67 & 1:33:48.27 & 30:33:15.7 & 22.05 & 2 & $-14.412$ & 1.246 & 0.028 & 1 &$-136.2$ & 7.9&&\\
  68 & 1:33:48.57 & 30:35:47.9 & 21.65 & 2 & $-14.548$ & 2.475 & 0.055 & 1 &$-95.5$ & 1.7 & 62&\\
  69 & 1:33:49.36 & 30:32:28.4 & 24.23 & 2 & $-14.381$ & 0.156 & 0.009 & 1 &$-119.5$ & 7.8&&\\
  70 & 1:33:49.41 & 30:32:06.6 & 21.73 & 2 & $-14.718$ & 3.398 & 0.088 & 2 &$-159.5$ & 2.9 & 65&\\
  71 & 1:33:50.02 & 30:14:25.2 & 21.25 & 1 & $-14.444$ & 2.818 & 0.056 & 2 &$-103.2$ & 4.0 & 63&\\
  72 & 1:33:50.82 & 30:18:43.8 & 23.90 & 1 & $-14.815$ & 0.575 & 0.029 & 3 &$-93.2$ & 6.2 & 64&\\
  73 & 1:33:51.06 & 30:45:38.8 & 21.28 & 2 & $-14.361$ & 2.247 & 0.050 & 3 &$-253.2$ & 2.1 & 95&\\
  74 & 1:33:52.05 & 30:48:56.3 & 23.63 & 1 & $-14.927$ & 0.954 & 0.072 & 0 &\nodata  & &&O\\
  75 & 1:33:52.64 & 30:16:54.2 & 21.05 & 1 & $-14.340$ & 2.642 & 0.045 & 2 &$-112.4$ & 4.7 & 67&\\
  76 & 1:33:52.77 & 30:37:38.9 & 20.78 & 2 & $-14.279$ & 2.948 & 0.045 & 2 &$-157.4$ & 1.0 & 68&O\\
  77 & 1:33:52.78 & 30:44:32.7 & 24.04 & 2 & $-15.304$ & 1.556 & 0.171 & 3 &$-229.6$ & 3.4 & 88&\\
  78 & 1:33:54.68 & 30:36:05.7 & 21.59 & 2 & $-14.206$ & 1.186 & 0.019 & 1 &$-147.4$ & 6.0 & 69&\\
  79 & 1:33:54.93 & 30:47:13.4 & 23.42 & 2 & $-15.170$ & 2.023 & 0.141 & 2 &$-240.6$ & 1.7&&\\
  80 & 1:33:55.58 & 30:16:02.9 & 23.94 & 1 & $-14.933$ & 0.728 & 0.033 & 1 &$-97.2$  & 8.0 & 71&O\\
  81 & 1:33:56.22 & 30:41:57.0 & 23.60 & 2 & $-15.057$ & 1.324 & 0.084 & 1 &$-157.1$ & 3.4&&H, O\\
  82 & 1:33:56.97 & 30:54:12.5 & 21.21 & 2 & $-14.451$ & 2.970 & 0.053 & 1 &$-260.8$ & 2.2 & 72&\\
  83 & 1:33:57.18 & 30:36:47.6 & 22.09 & 2 & $-14.871$ & 3.471 & 0.096 & 1 &$-134.7$ & 1.9 & 73&\\
  84 & 1:33:57.55 & 30:42:50.0 & 23.05 & 2 & $-15.145$ & 2.679 & 0.125 & 2 &$-230.6$ & 1.5&&\\
  85 & 1:33:58.38 & 30:55:05.2 & 22.91 & 2 & $-14.812$ & 1.415 & 0.072 & 1 &$-247.5$ & 3.6&&\\
  86 & 1:33:59.97 & 30:40:28.7 & 21.62 & 2 & $-14.627$ & 3.039 & 0.096 & 2 &$-212.8$ & 2.0 & 74&\\
  87 & 1:34:00.93 & 30:37:08.1 & 22.93 & 2 & $-15.668$ & 9.972 & 1.821 & 1 &$-156.4$ & 1.4&&\\
  88 & 1:34:01.12 & 30:50:27.3 & 20.95 & 2 & $-14.270$ & 2.484 & 0.031 & 2 &$-248.1$ & 2.4 & 75&O\\
  89 & 1:34:02.04 & 30:50:41.3 & 23.37 & 2 & $-14.741$ & 0.788 & 0.025 & 3 &$-253.6$ & 7.9 & 76&\\
  90 & 1:34:02.57 & 30:40:00.5 & 22.04 & 3 & $-14.782$ & 2.959 & 0.069 & 2 &$-188.3$ & 0.6 & 84&\\
  91 & 1:34:02.61 & 30:23:25.5 & 24.24 & 2 & $-15.551$ & 2.278 & 0.203 & 2 &$-161.2$ & 7.5 & 77&\\
  92 & 1:34:02.77 & 30:38:33.3 & 21.29 & 3 & $-14.328$ & 2.064 & 0.065 & 1 &$-192.3$ & 0.8&&O\\
  93 & 1:34:03.05 & 30:38:13.4 & 23.54 & 3 & $-15.014$ & 1.266 & 0.060 & 2 &$-190.8$ & 4.3&&\\
  94 & 1:34:03.54 & 30:39:15.6 & 21.25 & 3 & $-14.465$ & 2.930 & 0.050 & 2 &$-169.6$ & 2.5 & 79&O\\
  95 & 1:34:03.64 & 30:52:39.7 & 22.52 & 2 & $-14.910$ & 2.544 & 0.078 & 2 &$-258.6$ & 0.7 & 80&\\
  96 & 1:34:03.65 & 30:45:58.9 & 23.20 & 2 & $-14.484$ & 0.508 & 0.019 & 2 &$-205.8$ & 5.6 & 113&\\
  97 & 1:34:05.54 & 31:12:30.1 & 25.17 & 1 & $-15.752$ & 1.549 & 0.232 & 2 &\nodata  & &&\\
  98 & 1:34:05.82 & 30:36:13.9 & 24.14 & 2 & $-15.111$ & 0.910 & 0.072 & 1 &$-149.9$ & 8.8&&\\
  99 & 1:34:06.56 & 30:48:23.3 & 22.05 & 2 & $-14.724$ & 2.563 & 0.051 & 3 &$-263.2$ & 1.3 & 83&\\
 100 & 1:34:06.76 & 31:00:29.3 & 23.66 & 1 & $-15.246$ & 1.941 & 0.130 & 2 &$-279.8$ & 2.9 & 81&\\
 101 & 1:34:08.65 & 30:43:33.5 & 23.16 & 2 & $-14.495$ & 0.544 & 0.025 & 3 &$-213.7$ & 5.6 & 106&\\
 102 & 1:34:09.07 & 30:49:07.6 & 23.71 & 2 & $-14.784$ & 0.639 & 0.035 & 2 &$-239.5$ & 6.5 & 85&\\
 103 & 1:34:09.57 & 30:36:24.8 & 22.32 & 2 & $-14.720$ & 1.966 & 0.056 & 2 &$-173.5$ & 2.2&&\\
 104 & 1:34:11.74 & 31:07:31.4 & 22.84 & 1 & $-14.431$ & 0.627 & 0.023 & 1 &$-223.9$ & 7.0 & 86&\\
 105 & 1:34:11.91 & 30:45:00.9 & 21.32 & 2 & $-14.522$ & 3.145 & 0.054 & 1 &$-235.0$ & 2.1 & 119&\\
 106 & 1:34:12.15 & 30:28:57.1 & 23.86 & 2 & $-14.848$ & 0.641 & 0.030 & 0 &$-101.6$ & 11.3&&O\\
 107 & 1:34:13.34 & 30:45:01.2 & 21.00 & 2 & $-14.403$ & 3.227 & 0.052 & 1 &$-241.9$ & 1.8 & 111&\\
 108 & 1:34:13.40 & 30:33:50.7 & 21.21 & 2 & $-14.458$ & 3.006 & 0.046 & 2 &$-115.9$ & 1.5&&O\\
 109 & 1:34:13.98 & 30:22:36.5 & 20.63 & 1 & $-14.222$ & 2.981 & 0.067 & 2 &$-127.8$ & 1.6 & 91&O\\
 110 & 1:34:14.23 & 30:17:03.3 & 22.27 & 1 & $-14.483$ & 1.196 & 0.031 & 2 &$-125.1$ & 2.4 & 92&\\
 111 & 1:34:14.79 & 30:31:49.6 & 21.39 & 2 & $-14.410$ & 2.285 & 0.036 & 2 &$-159.0$ & 3.0 & 94&\\
 112 & 1:34:15.15 & 30:44:57.4 & 22.67 & 2 & $-14.244$ & 0.480 & 0.009 & 3 &$-211.0$ &12.1 & 123&\\
 113 & 1:34:15.48 & 30:32:20.3 & 20.96 & 2 & $-14.389$ & 3.221 & 0.053 & 1 &$-153.5$ & 1.1 & 96&\\
 114 & 1:34:15.72 & 31:08:12.5 & 21.72 & 1 & $-14.648$ & 2.903 & 0.082 & 1 &$-276.1$ & 2.9 & 89&\\
 115 & 1:34:15.88 & 30:24:54.5 & 21.47 & 2 & $-14.499$ & 2.597 & 0.038 & 2 &$-160.4$ & 1.0 & 93&\\
 116 & 1:34:16.61 & 30:41:10.4 & 22.01 & 2 & $-14.817$ & 3.272 & 0.076 & 2 &$-225.8$ & 2.1 & 105&\\
 117 & 1:34:18.54 & 30:58:30.5 & 21.90 & 1 & $-14.696$ & 2.752 & 0.074 & 2 &$-287.0$ & 2.6 & 97&\\
 118 & 1:34:19.21 & 30:44:57.0 & 21.13 & 2 & $-14.417$ & 2.940 & 0.055 & 2 &$-257.2$ & 1.9&&\\
 119 & 1:34:20.18 & 30:51:25.6 & 23.71 & 2 & $-15.070$ & 1.225 & 0.059 & 3 &$-265.6$ & 4.0 & 98&\\
 120 & 1:34:20.50 & 31:10:52.0 & 22.57 & 1 & $-14.370$ & 0.701 & 0.022 & 1 &\nodata  & &&\\
 121 & 1:34:22.92 & 31:10:38.6 & 25.91 & 1 & $-16.085$ & 1.686 & 0.442 & 1 &\nodata  & &&\\
 122 & 1:34:22.96 & 30:59:32.9 & 24.07 & 1 & $-15.120$ & 0.988 & 0.067 & 1 &$-242.7$ & 12.7 & 100&\\
 123 & 1:34:23.01 & 30:42:06.0 & 23.84 & 2 & $-14.931$ & 0.794 & 0.040 & 2 &$-216.1$ & 6.3 & 116&\\
 124 & 1:34:23.26 & 30:46:27.5 & 21.45 & 2 & $-14.529$ & 2.842 & 0.049 & 3 &$-259.7$ & 1.3 & 134&\\
 125 & 1:34:24.26 & 30:27:54.4 & 20.98 & 2 & $-14.401$ & 3.263 & 0.041 & 1 &$-135.4$ & 1.4 & 101&O\\
 126 & 1:34:25.09 & 30:39:39.7 & 22.27 & 3 & $-14.624$ & 1.652 & 0.033 & 1 &$-240.7$ & 1.5 & 103&\\
 127 & 1:34:25.53 & 30:17:30.0 & 25.16 & 1 & $-15.568$ & 1.017 & 0.123 & 3 &$-145.7$ & 11.6&&\\
 128 & 1:34:25.55 & 30:40:10.9 & 21.23 & 3 & $-14.429$ & 2.764 & 0.095 & 3 &$-175.3$ & 2.6 & 104&O\\
 129 & 1:34:25.83 & 31:07:44.5 & 24.28 & 1 & $-14.895$ & 0.486 & 0.032 & 1 &$-252.1$ & 9.6 & 102&\\
 130 & 1:34:27.71 & 30:42:32.5 & 22.25 & 2 & $-14.598$ & 1.585 & 0.044 & 3 &$-254.7$ & 2.7 & 126&\\
 131 & 1:34:28.00 & 30:47:45.8 & 25.00 & 2 & $-15.721$ & 1.672 & 0.255 & 4 &$-237.7$ & 9.0&&\\
 132 & 1:34:30.57 & 30:38:27.3 & 22.05 & 3 & $-14.524$ & 1.607 & 0.032 & 2 &$-154.3$ & 1.3 & 109&\\
 133 & 1:34:31.49 & 31:05:24.0 & 22.37 & 1 & $-14.507$ & 1.158 & 0.035 & 1 &$-252.3$ & 1.7 & 108&\\
 134 & 1:34:31.52 & 31:06:51.3 & 22.33 & 1 & $-14.940$ & 3.242 & 0.145 & 2 &$-262.7$ & 3.2 & 107&\\
 135 & 1:34:32.84 & 30:41:10.6 & 21.49 & 2 & $-14.376$ & 1.916 & 0.035 & 2 &$-188.4$ & 1.0 & 125&\\
 136 & 1:34:33.42 & 30:39:23.7 & 21.95 & 2 & $-14.700$ & 2.657 & 0.060 & 2 &$-208.2$ & 2.2 & 118&\\
 137 & 1:34:37.24 & 30:29:14.3 & 24.05 & 1 & $-15.321$ & 1.605 & 0.120 & 1 &$-145.9$ & 4.7&&\\
 138 & 1:34:37.50 & 30:40:12.4 & 24.38 & 2 & $-15.255$ & 1.020 & 0.062 & 3 &$-205.8$ & 4.3 & 129&\\
 139 & 1:34:37.89 & 30:27:00.7 & 21.57 & 2 & $-14.432$ & 2.038 & 0.038 & 2 &$-116.8$ & 1.4&&\\
 140 & 1:34:40.70 & 30:52:08.6 & 22.84 & 2 & $-14.405$ & 0.594 & 0.015 & 3 &$-274.0$ & 5.9 & 117&\\
 141 & 1:34:41.96 & 30:56:49.7 & 22.41 & 2 & $-14.642$ & 1.512 & 0.029 & 3 &$-239.4$ & 1.4 & 121&\\
 142 & 1:34:43.57 & 31:06:10.7 & 22.51 & 1 & $-14.506$ & 1.012 & 0.033 & 3 &$-251.9$ & 4.6 & 120&\\
 143 & 1:34:44.72 & 30:49:36.3 & 22.11 & 2 & $-14.160$ & 0.660 & 0.012 & 2 &$-250.0$ & 8.9 & 124&\\
 144 & 1:34:45.50 & 31:01:11.7 & 25.81 & 1 & $-16.095$ & 1.891 & 0.363 & 2 &\nodata  & &&\\
 145 & 1:34:46.18 & 30:44:41.7 & 22.50 & 2 & $-14.960$ & 2.900 & 0.122 & 0 &$-226.8$ & 26.8&&O\\
 146 & 1:34:47.28 & 30:27:27.5 & 24.09 & 1 & $-14.617$ & 0.307 & 0.021 & 2 &$-160.6$ &16.9&&\\
 147 & 1:34:58.18 & 31:06:47.5 & 22.51 & 1 & $-14.634$ & 1.355 & 0.052 & 3 &$-256.1$ & 3.4&&\\
 148 & 1:35:04.92 & 30:58:42.1 & 22.54 & 1 & $-14.802$ & 1.943 & 0.068 & 3 &$-221.0$ & 1.6&&\\
 149 & 1:35:08.63 & 30:49:32.8 & 21.13 & 2 & $-14.435$ & 3.062 & 0.056 & 1 &$-213.9$ & 3.7&&\\
 150 & 1:35:09.19 & 30:51:36.6 & 21.94 & 2 & $-14.752$ & 3.012 & 0.070 & 3 &$-255.0$ & 1.9&&\\
 151 & 1:35:10.50 & 31:05:14.9 & 24.41 & 1 & $-15.415$ & 1.430 & 0.128 & 1 &\nodata  & &&\\
 152 & 1:35:13.63 & 31:00:48.7 & 21.61 & 1 & $-14.467$ & 2.120 & 0.075 & 0 &\nodata  & &&O\\
\enddata
\tablenotetext{a}{Number of photometric measurements of the PN}
\tablenotetext{b}{H$\alpha$+[N~II] flux in units erg cm$^{-2}$ s$^{-1}$}
\tablenotetext{c}{Number of HYDRA setups in which PN was targeted (Oct 2002)}
\tablenotetext{d}{ID number from \citet{magrini33b} }
\tablecomments{O -- object also observed in Jan. 2003 observing 
run; NT -- object not targeted spectroscopically; H -- possible halo PN}
\label{tab2}
\end{deluxetable}

\clearpage
\begin{deluxetable}{ccccccc}
\tablewidth{0pt} 
\tablecaption{Mean Photometric Errors} 
\tablehead{
 \multicolumn{3}{c}{[O~III] Photometry}  &\colhead{\hskip20pt}
&\multicolumn{3}{c}{H$\alpha$ Photometry} \\
\colhead{$m_{5007}$}  &\colhead{$\sigma$(mag)}  &\colhead{N$_{\rm obj}$}
&&\colhead{$\log$ Flux} &\colhead{$\sigma$(mag)} &\colhead{N$_{\rm obj}$}}
\startdata
20.75 & 0.011 &  9 &&$-14.2$ &0.012  &10 \\
21.25 & 0.012 & 21 &&$-14.4$ &0.018  &40 \\
21.75 & 0.016 & 16 &&$-14.6$ &0.022  &24 \\
22.25 & 0.018 & 24 &&$-14.8$ &0.026  &26 \\
22.75 & 0.027 & 24 &&$-15.0$ &0.035  &13 \\
23.25 & 0.034 & 14 &&$-15.2$ &0.051  &12 \\
23.75 & 0.049 & 19 &&$-15.4$ &0.072  & 9 \\
24.25 & 0.065 & 14 &&$-15.6$ &0.100  & 6 \\
24.75 & 0.079 &  3 &&$-15.8$ &0.133  & 8 \\
\enddata
\label{tab3}
\end{deluxetable}

\begin{deluxetable}{cccl}
\tablewidth{0pt} \tabletypesize{\footnotesize}
\tablecaption{Unconfirmed \citet{magrini33b} Objects}
\tablehead{ 
\colhead{ID} &\colhead{$\alpha$ (2000)} &\colhead{$\delta$ (2000)} 
&\colhead{Reason}}
\startdata
  5 &1:32:44.04 &30:22:04.5  &Excess flux in $V$ and $B$\\
  8 &1:32:48.66 &30:25:53.2  &Excess flux in $V$ and $B$\\
 10 &1:32:54.20 &30:37:29.7  &Excess flux in $V$\\
 11 &1:33:05.74 &30:14:23.3  &Non-stellar in [O~III]; excess flux in $V$\\
 20 &1:33:09.61 &30:29:13.6  &Excess flux in $V$\\
 21 &1:33:08.43 &30:21:07.2  &Nothing at this location; non-stellar 
object at \citet{magrini33a} position\\
 23 &1:33:13.58 &30:22:36.3 &Non-stellar in $V$; excess flux in $B$\\
 24 &1:33:12.25 &30:30:49.1 &Excess flux in $V$\\
 34 &1:33:24.96 &30:38:48.7 &Non-stellar in [O~III]\\
 44 &1:33:33.59 &30:39:24.0 &Non-stellar in H$\alpha$\\
 47 &1:34:25.55 &30:40:10.7 &Coordinates identical to Magrini 104\\
 50 &1:33:39.53 &30:55:47.1 &Excess flux in $V$\\
 52 &1:33:37.37 &30:40:54.8 &Non-stellar in H$\alpha$\\
 53 &1:33:41.73 &30:37:24.1 &Excess flux in $V$\\
 55 &1:33:41.88 &30:08:31.3 &Excess flux in $V$\\
 56 &1:33:42.80 &30:54:04.3 &Non-stellar in $V$; excess flux in $V$x\\
 66 &1:33:50.88 &30:37:12.3 &Excess flux in $V$\\
 70 &1:33:54.97 &30:37:44.8 &Non-stellar in H$\alpha$\\
 78 &1:34:02.59 &30:58:10.2 &Non-stellar in [O~III]; excess flux in $V$\\
 87 &1:34:12.05 &30:39:10.1 &Excess flux in $V$\\
 90 &1:34:12.82 &30:47:18.1 &Non-stellar in H$\alpha$\\
 99 &1:34:20.61 &30:16:47.5 &Non-stellar in H$\alpha$\\
114 &1:34:16.39 &30:43:26.5 &Non-stellar in H$\alpha$; excess flux in $V$\\
115 &1:34:09.90 &30:45:07.8 &Non-stellar in [O~III]\\
122 &1:33:37.89 &30:39:34.8 &Non-stellar in [O~III] and $V$; excess flux in $V$\\
127 &1:34:47.13 &30:59:36.9 &Non-stellar in H$\alpha$; excess flux in $V$\\
128 &1:34:48.86 &31:05:14.8 &Non-stellar in H$\alpha$; excess flux in $V$\\
131 &1:34:51.19 &30:59:34.8 &Non-stellar in H$\alpha$; excess flux in $V$\\
132 &1:34:49.20 &30:37:27.7 &Excess flux in $V$ and $B$\\
133 &1:34:35.30 &30:43:56.5 &Excess flux in $V$\\
\enddata
\label{tab4}
\end{deluxetable}

\clearpage
\begin{deluxetable}{cccc}
\tablewidth{0pt} 
\tablecaption{Setup Velocity Differences} \tablehead{ \colhead{Setups} &
\colhead{$\overline{\Delta v}$} & \colhead{$\sigma_{\rm pair}$} &
\colhead{$N$}}
\startdata
1 $-$ 2 &  $-0.9$ &   1.0 & 5\\
1 $-$ 3 &  $-1.5$ &   2.7 & 8\\
1 $-$ 4 &  $-3.7$ &   4.9 & 22\\
1 $-$ 5 &  $+0.7$ &   3.3 & 22\\
1 $-$ 6 &  $-0.3$ &   7.2 & 17\\
2 $-$ 3 &  $+0.8$ &   8.6 & 18\\
2 $-$ 4 &  $-7.9$ &   6.4 &  6\\
2 $-$ 5 &  $-3.2$ &   6.4 & 19\\
2 $-$ 6 &  $-0.4$ &   5.3 & 12\\
3 $-$ 4 &  $-6.0$ &   4.9 & 13\\
3 $-$ 5 &  $-5.6$ &   5.8 &  9\\
3 $-$ 6 &  $-2.6$ &   4.8 & 16\\
4 $-$ 5 &  $+2.6$ &   3.3 & 10\\
4 $-$ 6 &  $+1.0$ &   6.3 & 21\\
5 $-$ 6 &  $+1.6$ &   7.7 & 11\\
\enddata
\label{tab5}
\end{deluxetable}

\begin{deluxetable}{ccccc}
\tablewidth{0pt}
\tablecaption{M33 Disk Mass and Mass-to-Light Ratios}
\tablehead{ 
\colhead{Radius} &\colhead{Surface Mass} & & &(Stellar) \\
\colhead{(kpc)}  &\colhead{($M_{\odot}$~pc$^{-2}$)} &\colhead{$M/L_K$} 
&\colhead{$M/L_V$} &\colhead{$M/L_V$} }
\startdata
1.2 &$79.2^{+49.3}_{-39.4}$ &$0.22^{+0.14}_{-0.11}$  &$0.52^{+0.33}_{-0.26}$
&$0.25^{+0.19}_{-0.15}$ \\[6pt]
2.1 &$54.1^{+31.4}_{-24.5}$ &$0.25^{+0.15}_{-0.11}$  &$0.50^{+0.29}_{-0.22}$
&$0.26^{+0.19}_{-0.15}$ \\[6pt]
2.7 &$82.3^{+33.3}_{-28.5}$ &$0.59^{+0.24}_{-0.21}$  &$1.00^{+0.41}_{-0.35}$
&$0.63^{+0.29}_{-0.25}$ \\[6pt]
3.4 &$51.8^{+21.3}_{-18.1}$ &$0.60^{+0.25}_{-0.21}$  &$0.85^{+0.35}_{-0.30}$
&$0.55^{+0.27}_{-0.23}$ \\[6pt]
4.5 &$31.6^{+13.3}_{-11.2}$ &$0.74^{+0.31}_{-0.26}$  &$0.82^{+0.34}_{-0.29}$
&$0.50^{+0.29}_{-0.25}$ \\[6pt]
6.0 &$31.0^{+9.3}_{-8.3}$  &$1.79^{+0.54}_{-0.48}$   &$1.42^{+0.43}_{-0.38}$
&$0.99^{+0.39}_{-0.34}$ \\[6pt]
7.1 &$19.2^{+6.1}_{-5.4}$  &$2.34^{+0.74}_{-0.66}$   &$1.42^{+0.45}_{-0.40}$
&$0.98^{+0.42}_{-0.37}$ \\[6pt]
8.5 &$16.3^{+5.0}_{-4.5}$  &$4.68^{+1.44}_{-1.29}$   &$2.07^{+0.64}_{-0.57}$ 
&$1.52^{+0.61}_{-0.55}$ \\
\enddata
\label{tab6}
\end{deluxetable}

\clearpage

\clearpage

\clearpage

\clearpage

\clearpage

\clearpage

\clearpage

\clearpage

\clearpage

\clearpage

\clearpage

\clearpage

\clearpage


\clearpage

\begin{thebibliography}{}

\bibitem[Araki(1985)]{araki} Araki, S. 1985, Ph.D. thesis, Massachusetts
Institute of Technology
\bibitem[Ashman(1992)]{ashman} Ashman, K.M. 1992, \pasp, 104, 1109 
\bibitem[Bell \& de Jong(2000)]{BdJ00} Bell, E.F., \& de Jong, R.S. 2000,
\apj, 312, 497
\bibitem[Bell \& de Jong(2001)]{BdJ01} Bell, E.F., \& de Jong, R.S. 2001,
\apj, 550, 212
\bibitem[Berkhuijsen(1983)]{berkhuijsen} Berkhuijsen, E.M. 1983, \aap, 127, 395
\bibitem[Bevington(1969)]{bevington} Bevington, P.R. 1969, Data Reduction and
Error Analysis for the Physical Sciences (New York: McGraw-Hill)
\bibitem[Bienaym\'e(1999)]{bienayme} Bienaym\'e, O. 1999, \aap, 341, 86
\bibitem[Binney \& Tremaine(1987)]{bt} Binney, J., \& Tremaine, S. 1987,
Galactic Dynamics, (Princeton: Princeton Univ.~Press)
\bibitem[Bizyaev \& Mitronova(2002)]{bm02} Bizyaev, D., \& Mitronova, S. 
2002, \aap, 389, 795
\bibitem[Borriello \& Salucci(2001)]{bs01} Borriello, A., \& Salucci, P.
2001, \mnras, 323, 285
\bibitem[Bothun(1992)]{bothun} Bothun, G.D. 1992, \aj, 103, 104
\bibitem[Bottema(1993)]{bottema}  Bottema, R. 1993, \aap, 275, 16
\bibitem[Bottema, van der Kruit, \& Freeman(1987)]{n5170} Bottema, R.,
van der Kruit, P.C., \& Freeman, K.C. 1987, \aap, 178, 77
\bibitem[Brandt(1960)]{brandt} Brandt, J.C. 1960, \apj, 131, 293
\bibitem[Brandt(1965)]{brandt65} Brandt, J.C. 1965, \mnras, 129, 309
\bibitem[Cardelli, Clayton, \& Mathis(1989)]{ccm} Cardelli, J.A.,
Clayton, G.C., \& Mathis, J.S. 1989, \apj, 345, 245
\bibitem[Carlberg(1987)]{carlberg} Carlberg, R.G. 1987, \apj, 322, 59
\bibitem[Chen \etal(2001)]{chen} Chen, B., Stoughton, C., Smith, J.A.,
Uomoto, A., Pier, J.R., Yanny, B., Ivez\'ic, \u Z, York, D.G., 
Anderson, J.E., Annis, J., Brinkmann, J., Csabai, I., Fukugita, M., 
Hindsley, R., Lupton, R., \& Munn, J.A. 2001, \apj, 553, 184
\bibitem[Ciardullo(2003)]{chile} Ciardullo, R. 2003 in Lecture Notes
in Physics: Stellar Candles for the Extragalactic Distance Scale, ed.~W.
Gieren \& D. Alloin (Heidelberg: Springer-Verlag), 243

\bibitem[Ciardullo \etal(1987)]{cfnjs} Ciardullo, R., Ford, H.C., Neill, J.D.,
Jacoby, G.H., \& Shafter, A.W. 1987, \apj, 318, 520

\bibitem[Ciardullo \etal(2002)]{p12} Ciardullo, R., Feldmeier, J.J.,
Jacoby, G.H., de Naray, R.K., Laychak, M.B., \& Durrell, P.R. 2002,
\apj, 577, 31

\bibitem[Ciardullo \& Jacoby(1999)]{cj99} Ciardullo, R., \& Jacoby, G.H. 1999,
\apj, 515, 191

\bibitem[Ciardullo \etal(1989)]{p2} Ciardullo, R., Jacoby, G.H., Ford, H.C., \&
Neill, J.D. 1989, \apj, 339, 53

\bibitem[Ciardullo, Jacoby, \& Harris(1991)]{p7} Ciardullo, R., Jacoby, G.H.,
\& Harris, W.E. 1991, \apj, 383, 487

\bibitem[Combes(2002)]{combes} Combes, F. 2002, New Astronomy Reviews, 46,
755

\bibitem[Corbelli(2003)]{c03} Corbelli, E. 2003, \mnras, 342, 199

\bibitem[Corbelli \& Salucci(2000)]{cs00} Corbelli, E., \& Salucci, P.
2000, \mnras, 311, 441

\bibitem[Corbelli \& Schneider(1997)]{cs97} Corbelli, E., \& 
Schneider, S.E. 1997, \apj, 479, 244 

\bibitem[Corradi \& Schwarz(1995)]{cs95} Corradi, R.L.M., \& Schwarz, H.E.
1995, \aap, 293, 871

\bibitem[Dalcanton \& Bernstein(2002)]{dalcanton} Dalcanton, J.J., \&
Bernstein, R.A. 2002, \aj, 124, 1328

\bibitem[Dav\'e \etal(2001)]{dave} Dav\'e, R., Spergel, D.N., Steinhardt, P.J.,
\& Wandelt, B.D. 2001, \apj, 547, 574

\bibitem[de Blok, Bosma, \& McGaugh(2003)]{dBBMc} de Blok, W.J.G., Bosma, A., 
\& McGaugh, S. 2003, \mnras, 340, 657

\bibitem[de Grijs \& Peletier(1997)]{dG97} de Grijs, R., \& Peletier, R.F.
1997, \aap, 320, L21

\bibitem[de Grijs \& van der Kruit(1996)]{dG96} de Grijs, R., \& van der 
Kruit, P.C. 1996, \aaps, 117, 19

\bibitem[de Vaucouleurs(1959)]{deV59} de Vaucouleurs, G. 1959, \apj, 130, 728

\bibitem[de Vaucouleurs \etal(1991)]{RC3} de Vaucouleurs, G., de Vaucouleurs,
A., Corwin, H.G. Jr., Buta, R.J., Paturel, G., \& Fouqu\'e, P. 1991,
Third Reference Catalog of Bright Galaxies, (New York: Springer-Verlag)

\bibitem[Dehnen \& Binney(1998)]{db98} Dehnen, W., \& Binney, J.J. 1998,
\mnras, 298, 387

\bibitem[Dopita, Jacoby, \& Vassiliadis(1992)]{djv92} Dopita, M.A., Jacoby, 
G.H., \& Vassiliadis, E. 1992, \apj, 389, 27 

\bibitem[Dopita \etal(1997)]{dopita} Dopita, M.A., Vassiliadis, E., Wood, 
P.R., Meatheringham, S.J., Harrington, J.P., Bohlin, R.C., Ford, H.C., 
Stecher, T.P., \& Maran, S.P. 1997, \apj, 474, 188

\bibitem[Eggen, Lynden-Bell, \& Sandage(1962)]{els} Eggen, O.J., 
Lynden-Bell, D., \& Sandage, A.R. 1962, \apj, 136, 748

\bibitem[Faber \& Gallagher(1979)]{fg79} Faber, S.M., \& Gallagher, J.S. 1979,
\araa, 17, 135

\bibitem[Fall \& Efstathiou(1980)]{fe80} Fall, S.M., \& Efstathiou, G.
1980, \mnras, 193, 189

\bibitem[Freedman \etal(2001)]{keyfinal} Freedman, W.L., Madore, B.F.,
Gibson, B.K., Ferrarese, L., Kelson, D.D., Sakai, S., Mould, J.R.,
Kennicutt, R.C. Jr., Ford, H.C., Graham, J.A., Huchra, J.P., Hughes,
S.M.G., Illingworth, G.D., Macri, L.M., \& Stetson, P.B. 2001, \apj, 553, 47

\bibitem[Freedman, Wilson, \& Madore(1991)]{freed91} Freedman, W.L.,
Wilson, C.D., \& Madore, B.F. 1991, \apj, 372, 455

\bibitem[Fry \etal(1999)]{fry99} Fry, A.M., Morrison, H.L., Harding, P., \&
Boroson, T.A. 1999, \aj, 118, 1209

\bibitem[Fuchs \& Wielen(1987)]{fw87} Fuchs, B., \& Wielen, R. 1987,
in NATO ASIC Proc.~207, The Galaxy, ed.~G. Gilmore \& B. Carswell
(Dordrecht: Reidel), 375

\bibitem[Garnett \etal(1997)]{garnett} Garnett, D.R., Shields, G.A.,
Skillman, E.D., Sagan, S.P., \& Dufour, R.J. 1997, \apj, 489, 63

\bibitem[Gebhardt \etal(2001)]{gebhardt} Gebhardt, K., Lauer, T.R., 
Kormendy, J., Pinkney, J., Bower, G.A., Green, R., Gull, T., Hutchings, J.B.,
Kaiser, M.E., Nelson, C.H., Richstone, D., \& Weistrop, D. 2001, \aj, 122,
2469

\bibitem[Gerssen, Kuijken, \& Merrifield(1997)]{n488} Gerssen, J.,
Kuijken, K., \& Merrifield, M.R. 1997, \mnras, 288, 618

\bibitem[Gerssen, Kuijken, \& Merrifield(2000)]{n2985} Gerssen, J.,
Kuijken, K., \& Merrifield, M.R. 2000, \mnras, 317, 545

\bibitem[Giraud(2000)]{g00} Giraud, E. 2000, \apj, 539, 155

\bibitem[Griv \etal(1999)]{griv} Griv, E., Rosenstein, B., Gedalin, M.,
\& Eichler, D. 1999, \aap, 347, 821

\bibitem[Guidoni, Messi, \& Natali(1981)]{guidoni} Guidoni, U., Messi, R., \&
Natali, G. 1981, \aap, 96, 215

\bibitem[Henize \& Westerlund(1963)]{hw63} Henize, K.G., \& Westerlund, B.E.
1963, \apj, 137, 747

\bibitem[Heyer \etal(2004)]{heyer} Heyer, M.H., Corbelli, E.,
Schneider, S.E., \& Young, J.S. 2004, \apj, 602, 723

\bibitem[Hippelein \etal(2003)]{hipp} Hippelein, H., Haas, M., Tuffs, R.J.,
Lemke, D., Stickel, M., Klaas, U., \& V\"olk, H.J. 2003, \aap, 407, 137

\bibitem[Hodge \etal(1999)]{hodge} Hodge, P.W., Balsley, J., Wyder, T.K., \&
Skelton, B.P. 1999, \pasp, 111, 685

\bibitem[Huchra, Vogeley, \& Geller(1999)]{hvg99} Huchra, J.P., Vogeley,
M.S., \& Geller, M.J. 1999, \apjs, 121, 287

\bibitem[Hui \etal(1993)]{hui93} Hui, X., Ford, H.C., Ciardullo, R., \&
Jacoby, G.H. 1993, \apj, 414, 463

\bibitem[Iben \& Laughlin(1989)]{iben} Iben, I., Jr., \& Laughlin, G. 1989,
\apj, 341, 312

\bibitem[Iben \& Renzini(1983)]{ir83} Iben, I., Jr., \& Renzini, A. 1983, 
\araa, 21, 271 

\bibitem[Israel \& Kennicutt(1980)]{israel} Israel, F.P., \& Kennicutt, R.C.
1980, \aplett, 21, 1

\bibitem[Jacoby(1989)]{p1} Jacoby, G.H. 1989, \apj, 339, 39

\bibitem[Jacoby(1997)]{jacoby97} Jacoby, G.H. 1997, in The Extragalactic
Distance Scale, Space Telescope Science Institute Series, ed.~M. Livio
(Cambridge: Cambridge University Press), 197 

\bibitem[Jacoby \etal(1992)]{mudville} Jacoby, G.H., Branch, D., Ciardullo, R.,
Davies, R.L., Harris, W.E., Pierce, M.J., Pritchet, C.J., Tonry, J.L., \&
Welch, D.L. 1992, \pasp, 104, 599

\bibitem[Jacoby \etal(1989)]{p3} Jacoby, G.H., Ciardullo, R., Ford, H.C., \&
Booth, J. 1989, \apj, 344, 70

\bibitem[Jacoby, Ciardullo, \& Harris(1996)]{p10} Jacoby, G.H., Ciardullo, R.,
\& Harris, W.E. 1996, \apj, 462, 1

\bibitem[Jacoby \& De Marco(2002)]{jd02} Jacoby, G.H., \& De Marco, O. 2002,
\aj, 123, 269

\bibitem[Jacoby, Quigley, \& Africano(1987)]{jqa} Jacoby, G.H., Quigley, R.J.,
\& Africano, J.L. 1987, \pasp, 99, 672

\bibitem[Jahrei\ss\ \& Wielen(1983)]{jw83} Jahrei\ss, H., \& Wielen, R. 1983,
IAU Colloq.~76, Nearby Stars and the Stellar Luminosity Function,
ed.~A.G.D. Philip \& A.R. Upgren (Schenectady: L. Davis Press), 277

\bibitem[Jenkins \& Binney(1990)]{jb90} Jenkins, A., \& Binney, J. 1990,
\mnras, 245, 305

\bibitem[Kaler \& Jacoby(1990)]{kj90} Kaler, J.B., \& Jacoby, G.H. 1990,
\apj, 362, 491

\bibitem[Kennicutt, Tamblyn, \& Congdon(1994)]{ktc} Kennicutt, R.C., Tamblyn, 
P., \& Congdon, C.E. 1994, \apj, 435, 22 

\bibitem[Kent(1986)]{kent86} Kent, S.M. 1986, \aj, 91, 1301

\bibitem[Kent(1987)]{kent} Kent, S.M. 1987, \aj, 94, 306

\bibitem[Khoperskov, Zasov, \& Tyurina(2003)]{khoperskov} Khoperskov, A.V.,
Zasov, A.V., \& Tyurina, N.V. 2003, Astronomy Reports, 47, 357

\bibitem[Kim \etal(2002)]{kim} Kim, M., Kim, E., Lee, M.G., Sarajedini, A.,
\& Geisler, D. 2002, \aj, 123, 244

\bibitem[Klypin \etal(2001)]{klypin} Klypin, A., Kravtsov, A.V., Bullock, J.S.,
\& Primack, J.R. 2001, \apj, 554, 903

\bibitem[Kregel, van der Kruit, \& de Grijs(2002)]{kregel} Kregel, M., 
van der Kruit, P.C., \& de Grijs, R. 2002, \mnras, 334, 646

\bibitem[Kurtz \& Mink(1998)]{rvsao}Kurtz, M.J., \& Mink, D.J. 1998,
\pasp, 110, 934

\bibitem[Maciel(1989)]{maciel} Maciel, W.J. 1989, in IAU Symp.~131,
Planetary Nebulae, ed.~S. Torres-Peimbert (Dordrecht: Kluwer), 73

\bibitem[Magrini \etal(2001)]{magrini33b} Magrini, L., Cardwell, A.,
Corradi, R.L.M., Mampaso, A., \& Perinotto, M. 2001, \aap, 367, 498

\bibitem[Magrini \etal(2000)]{magrini33a} Magrini, L., Corradi, R.L.M.,
Mampaso, A., \& Perinotto, M. 2000, \aap, 355, 713

\bibitem[Magrini \etal(2003)]{magrini03} Magrini, L., Perinotto, M.,
Corradi, R.L.M., \& Mampaso, A. 2003, \aap, 400, 511

\bibitem[Maraston(1998)]{maraston} Maraston, C. 1998, \mnras, 300, 872

\bibitem[Marigo \etal(2004)]{marigo} Marigo, P., Girardi, L., Weiss, A.,
Groenewegen, M.A.T., \& Chiosi, C. 2004, \aap, in press

\bibitem[Massey \etal(1996)]{massey96} Massey, P., Bianchi, L., 
Hutchings, J.B., \& Stecher, T.P. 1996, \apj, 469, 629

\bibitem[Massey \etal(2002)]{survey}Massey, P., Hodge, P.W., Holmes, S., 
Jacoby, J., King, N.L., Olsen, K., Smith, C., \& Saha, A. 2002, \baas, 34, 
1272

\bibitem[Merritt \& Sellwood(1994)]{ms94} Merritt, D., \& Sellwood, J.A.
1994, \apj, 425, 567

\bibitem[Mihalas \& Binney(1981)]{mb81} Mihalas, D., \& Binney, J. 1981
Galactic Astronomy (New York: Freeman)

\bibitem[Monet \etal(1998)]{monet} Monet, D., Bird, A., Canzian, B.,
Dahn, C., Guetter, H., Harris, H., Henden, A., Levine, S., Luginbuhl, C.,
Monet, A.K.B., Rhodes, A., Riepe, B., Sell, S., Stone, R., Vrba, F.,
\& Walker, R. 1998, PMM USNO-A2.0: A Catalogue of Astrometric Standards
(Washington, DC: US Naval Obs.)

\bibitem[Moore \etal(1998)]{moore} Moore, B., Governato, F., Quinn, T., 
Stadel, J., \& Lake, G.\ 1998, \apjl, 499, L5

\bibitem[Moriondo, Giovanardi, \& Hunt(1998)]{moriondo} Moriondo, G.,
Giovanardi, C., \& Hunt, L.K. 1998, \aap, 339, 409

\bibitem[Morosov(1980)]{moro1}  Morozov, A.G. 1980, SvA, 24, 391

\bibitem[Morosov(1981a)]{moro2} Morozov, A.G. 1980, SvA, 25, 19

\bibitem[Morosov(1981b)]{moro3} Morozov, A.G. 1980, SvA, 25, 421

\bibitem[Navarro, Frenk, \& White(1996)]{nfw} Navarro, J.F., Frenk, C.S.,
\& White, S.D.M. 1996, \apj, 462, 563

\bibitem[Palunas \& Williams(2000)]{palunas} Palunas, P., \& Williams, T.B.
2000, \aj, 120, 2884

\bibitem[Peimbert(1993)]{peimbert} Peimbert, M. 1993, in IAU Symp.~155,
Planetary Nebulae, ed.~R. Weinberger \& A. Acker (Dordrecht: Kluwer), 523

\bibitem[Petersen \& Gammelgaard(1997)]{pg97} Petersen, L., \& Gammelgaard, P.
1997, \aap, 323, 697

\bibitem[Pfenniger, Combes, \& Martinet(1994)]{pfenniger} Pfenniger, D.,
Combes, F., \& Martinet, L. 1994, \aap, 285, 79

\bibitem[Phillips(2003)]{phillips} Phillips, J.P. 2003, New Astronomy, 8, 29

\bibitem[Pottasch(1984)]{pot84} Pottasch, S.R. 1984, Planetary Nebulae,
(Dordrecht: Reidel), Chapter 5

\bibitem[Power \etal(2003)]{power} Power, C., Navarro, J.F., Jenkins, A.,
Frenk, C.S., White, S.D.M., Springel, V., Stadel, J., \& Quinn, T. 2003,
\mnras, 338, 14

\bibitem[Ratnatunga \& Upgren(1997)]{ku97} Ratnatunga, K.U., \& Upgren,
A.R. 1997, \apj, 476, 811

\bibitem[Regan \& Vogel(1994)]{rv94} Regan, M.W., \& Vogel, S.N. 1994, 
\apj, 434, 536

\bibitem[Rogstad, Wright, \& Lockhart(1976)]{rwl76} Rogstad, D.H., 
Wright, M.C.H., \& Lockhart, I.A. 1976, \apj, 204, 703 

\bibitem[Rubin(1997)]{rubin} Rubin, V. 1997, Bright Galaxies, Dark Matters,
(Woodbury, NY: American Institute of Physics)

\bibitem[Ryder \& Dopita(1994)]{ryder} Ryder, S.D., \& Dopita, M.A. 1994,
\apj, 430, 142

\bibitem[Sandage \& Humphreys(1980)]{sh80} Sandage, A., \& Humphreys, R.M.
1980, \apjl, 263, L1

\bibitem[Scalo(1986)]{scalo} Scalo, J.M. 1986, FCPh, 11, 1

\bibitem[Schlegel, Finkbeiner, \& Davis(1998)]{schlegel} Schlegel, D.J.,
Finkbeiner, D.P., \& Davis, M. 1998, \apj, 500, 525

\bibitem[Schwarzkopf \& Dettmar(2001)]{schwarz01} Schwarzkopf, U., \&
Dettmar, R.-J. 2001, \aap, 373, 402

\bibitem[Searle \& Zinn(1978)]{sz} Searle, L., \& Zinn, R. 1978, \apj,
225, 357

\bibitem[Sellwood(1996)]{sellwood} Sellwood, J.A. 1996, \apj, 473, 733

\bibitem[Shaver \etal(1983)]{shaver} Shaver, P.A., McGee, R.X., Newton, L.M.,
Danks, A.C., \& Pottasch, S.R. 1983, \mnras, 204, 53

\bibitem[Shaw \& Gilmore(1990)]{sg90} Shaw, M.A., \& Gilmore, G. 1990, 
\mnras, 242, 59

\bibitem[Sofue \etal(2003)]{sofue} Sofue, Y., Koda, J., Nakanishi, H., \&
Onodera, S. 2003, \pasj, 55, 59,

\bibitem[Spekkens, \& Giovanelli(2003)]{spekkens} Spekkens, K., \&
Giovanelli, R. 2003, in IAU Symp.~220, Dark Matter in Galaxies, ed.~S. Ryder,
D.J. Pisano, M. Walker, \& K. Freeman (San Francisco: ASP), in press

\bibitem[Steinmetz \& M\"uller(1994)]{sm94} Steinmetz, M., \& M\"uller, E.
1994, \aap, 281, 97

\bibitem[Stetson(1987)]{stet87}Stetson, P.B. 1987, \pasp, 99, 191

\bibitem[Stetson(1992)]{stet92}Stetson, P.B. 1992, in ASP Conf.~Ser.~25,
Astronomical Data Analysis Software and Systems I., ed.~D.M. Worral,
C. Biemesderfer, \& J. Barnes (San Francisco: ASP), 297

\bibitem[Stetson, Davis, \& Crabtree(1990)]{stet90} Stetson, P.B., Davis, L.E.,
\& Crabtree, D.R. 1990, in ASP~Conf.~Ser.~8, CCDs in Astronomy,
ed.~G.H. Jacoby (San Francisco: ASP), 289

\bibitem[Stone(1977)]{stone} Stone, R.P.S. 1977, \apj, 218, 767

\bibitem[Swaters \etal(2003)]{swaters} Swaters, R.A.,
Madore, B.F., van den Bosch, F.C., \& Balcells, M. 2003, \apj, 583, 732

\bibitem[Tiede, Sarajedini, \& Barker(2004)]{tiede} Tiede, G.P., 
Sarajedini, A., \& Barker, M.K. 2004, \aj, in press

\bibitem[Toomre(1964)]{toomre} Toomre, A. 1964, \apj, 139, 1217

\bibitem[Toomre(1966)]{toomre66} Toomre, A. 1966, in Notes on the 1966
Summer Study Program in Geophysical Fluid Dynamics at the Woods Hole
Oceanographic Institution (Woods Hole: Woods Hole Oceanographic Institute), 111

\bibitem[Udalski \etal(1999)]{udalski99} Udalski, A., Szyma\'nski, M.,
Kubiak, M., Pietrzy\'nski, G., Soszy\'nski, I., Wo\'zniak, P., \&
\.Zebru\'n, K. 1999, Act.~Astr., 49, 201

\bibitem[Valdes(1998)]{valdes} Valdes, F.G. 1998, in ASP~Conf.~Ser.~145,
Astronomical Data Analysis Software and Systems VII, ed.~R. Albrecht,
R.N. Hook, \& H.A. Bushhouse (San Francisco: ASP), 53

\bibitem[van der Kruit(1988)]{vdK88} van der Kruit, P.C. 1988, \aap, 192, 117

\bibitem[van der Kruit \& de Grijs(1999)]{vdKdG} van der Kruit, P.C., \&
de Grijs, R. 1999, \aap, 352, 129

\bibitem[van der Kruit \& Searle(1982)] {vdK82} van der Kruit, P.C., \& 
Searle, L. 1982, \aap, 110, 61

\bibitem[van der Kruit \etal(2001)]{ic5249} van der Kruit, P.C., 
Jim\'enez-Vicente, J., Kregel, M., \& Freeman, K.C. 2001, \aap, 379, 374

\bibitem[Vassiliadis \& Wood(1994)]{vw94} Vassiliadis, E., \& Wood, P.R.
1994, \apjs, 92, 125

\bibitem[Villumsen(1985)]{villumsen} Villumsen, J.V. 1985, \apj, 290, 75

\bibitem[Weidemann(2000)]{weidemann} Weidemann, V. 2000, \aap, 363, 647

\bibitem[Wielen(1977)]{wielen} Wielen, R. 1977, \aap, 60, 263

\bibitem[Zaritsky, Elston, \& Hill(1989)]{zar89} Zaritsky, D., Elston, R.,
\& Hill, J.M. 1989, \aj, 97, 97

\bibitem[Zijlstra \& Pottasch(1991)]{zp91} Zijlstra, A.A., \& Pottasch, S.R.
1991, \aap, 243, 478

\end{thebibliography}
\end{document}